\documentclass[11pt,a4paper,onecolumn]{article}
\usepackage[margin=22mm]{geometry}
\usepackage[utf8]{inputenc}
\usepackage[T1]{fontenc}
\usepackage{lmodern}
\usepackage{microtype}
\usepackage[table]{xcolor}
\usepackage{float}
\usepackage{subfig}
\usepackage{graphicx}
\usepackage{multirow}
\usepackage[super]{nth}
\usepackage{booktabs}
\usepackage{amsfonts,amsmath,amsthm,amssymb}
\usepackage{nicefrac}
\usepackage[square,comma,sort&compress,numbers]{natbib}
\usepackage{algorithm}
\usepackage{algorithmic}
\usepackage{adjustbox}
\usepackage[section]{placeins}
\AddToHook{cmd/subsection/before}{\FloatBarrier}
\usepackage[font=small,labelfont=bf]{caption}
\usepackage{authblk}
\usepackage{xurl}
\usepackage[hidelinks]{hyperref}
\hypersetup{pdftitle={BFLA: Block-Filtered Long-Context Attention Mechanism},pdfauthor={Chong Wu et al.}}
\newtheorem{bflaboundlemma}{Lemma}[section]
\newtheorem{bflaboundtheorem}[bflaboundlemma]{Theorem}
\newtheorem{bflaboundcorollary}[bflaboundlemma]{Corollary}
\theoremstyle{remark}

\makeatletter
\@addtoreset{equation}{myequation}
\makeatother

\title{BFLA: Block-Filtered Long-Context Attention Mechanism}

\author[1]{Chong~Wu\thanks{Corresponding Author}\thanks{Project Leader}\thanks{Equal Contribution}}
\author[2]{Brian~K~Chen\protect\footnotemark[3]}
\author[3]{Chuxu~Song\protect\footnotemark[3]}
\author[4]{Shezheng~Song\protect\footnotemark[3]}
\author[1]{Zhenan~Feng\protect\footnotemark[3]}
\author[4]{Renjie~Xu}
\author[1]{Houwang~Zhang}
\author[5]{Jiawang~Cao}
\author[6]{Maolin~Che\protect\footnotemark[1]}
\author[7]{Wenbo~Zhu}
\author[1]{Hong~Yan}
\affil[1]{City University of Hong Kong}
\affil[2]{National University of Singapore}
\affil[3]{Rutgers University}
\affil[4]{Individual researcher}
\affil[5]{Bytedance}
\affil[6]{Guizhou University}
\affil[7]{University of California, Berkeley}
\affil[ ]{\texttt{imroxaswc@gmail.com \& chncml@outlook.com}}
\date{}
\begin{document}
\maketitle

\begin{abstract}
We propose Block-Filtered Long-Context Attention (BFLA), a training-free sparse prefill method that combines deterministic Lightning-style block indexing, query-tile sharing, and length-adaptive Ratio-$K$ selection with block-sparse execution. BFLA evaluates attention using the original Q/K/V, requires no retraining or per-model parameter retuning, and integrates with vLLM while preserving native decoding. For fixed Q/K/V, we bound local attention-output error in terms of the omitted attention mass, indexer-score distortion, retention budget, and query-tile sharing. Evaluations on Gemma4-E4B-IT, Qwen3.6-27B, Qwen3.8-27B, and Qwen3.8-Flash-Next across LongBench-v2, RULER, InfiniteBench, and Terminal Bench 2.1 show BFLA matches or even outperforms dense baselines. BFLA shows up to approximately $4\times$ time-to-first-token and $3.72\times$ end-to-end request speedups over the corresponding dense-prefill baselines. Comparisons with state-of-the-art FlashPrefill V2 and Qwen Sparse Attention also demonstrate that BFLA has better performance and efficiency trade-offs. These findings highlight the value of BFLA for the acceleration of long-context serving. Project website: \url{https://github.com/Alicewithrabbit/BFLA}.
\end{abstract}

\section{Introduction}
The transformer architecture \citep{ATA} has become the dominant backbone for large language models (LLMs) \citep{Llama,GPT}. During prefill, dense attention compares queries with causally available keys, requiring arithmetic that grows quadratically with the sequence length $N$ for fixed head geometry.

Existing approaches to efficient attention can be broadly classified into three complementary directions: \textbf{(1) Sparse attention}, which exploits concentrated attention distributions by restricting computation to selected entries \citep{SPARSE3,SPARSE2,SPARSE5,xu2025xattention,spargeattn}; \textbf{(2) Linear attention}, which uses feature-map factorizations to achieve $O(N)$ complexity for fixed feature dimensions \citep{TRNN,PEF,COS,flatten}; and \textbf{(3) Hybrid attention}, which combines different attention mechanisms in a single unified architecture \citep{elfatt,NSA}.

Among these paradigms, sparse attention has a unique advantage: it preserves the original softmax attention mechanism and can directly approximate full attention output without altering the model weights. This property makes sparse attention methods a natural basis for training-free acceleration. However, not all sparse-attention methods are training-free: some learn the indexer or adapt the backbone to the sparse pattern. Static sparse methods use input-independent patterns, such as local windows, strided connections, or designated global tokens \citep{SPARSE2,SPARSE3,SPARSE5,Swin}, whereas dynamic sparse methods select tokens or blocks from the current input \citep{xu2025xattention,spargeattn,MInference,Quest}. Realizing dynamic sparsity efficiently requires both lightweight support selection and an implementation that avoids computing omitted entries. In long-context prefill, indexing and runtime overhead can otherwise offset the savings from sparse attention.

We present \textbf{Block-Filtered Long-Context Attention (BFLA)}, a training-free sparse prefill method that jointly designs compact indexing, support selection, and execution. A deterministic Lightning-style indexer derives compact query features and block-level key features from post-RoPE Q/K. Ratio-$K$ selection assigns a context-dependent budget with lower and upper limits $K_{\min}$ and $K_{\max}$, clipped to the visible length of each query tile. Query-tile sharing constructs one block list for multiple query vectors. BFLA computes attention over the selected blocks and integrates with vLLM for full and chunked prefill. Section~\ref{sec:theory} bounds local attention error through score distortion and retained attention mass, explaining when a larger budget can compensate for less accurate indexing. Our contributions are:

\begin{itemize}
    \item[(i)] \textbf{A training-free Lightning-style block indexer.}
    We construct compact query features and a block-level FP8 indexer key state from post-RoPE Q/K without learned indexer projections or model-weight modifications.

    \item[(ii)] \textbf{Length-adaptive Ratio-$K$ selection.}
    We pair the compact indexer with a retention budget that combines a minimum, proportional growth with context length, and a maximum selection limit, while incorporating mandatory blocks and protected full-support tail queries.

    \item[(iii)] \textbf{Theoretical support for the indexer--budget tradeoff.}
    We bound local attention error through the omitted attention mass and relate this mass to index-ranking distortion, the retention budget, and query-tile sharing, formalizing when a larger retained set can tolerate a coarser ranking.

    \item[(iv)] \textbf{Query-tile sharing with block-sparse execution.}
    We construct one shared block support per query tile and execute it in a block-sparse attention kernel. This couples support selection to K/V reuse and skips omitted blocks in the attention loop.

    \item[(v)] \textbf{An end-to-end vLLM implementation.}
    We integrate indexing, dynamic support selection, and block-sparse attention into vLLM while preserving native decoding, and evaluate one shared sparsity configuration across different models without per-model retuning.
\end{itemize}

\section{Related Work}

\subsection{Sparse attention}

\subsubsection{Static sparse attention}
Static sparse attention methods typically use predetermined, input-independent patterns to reduce the number of computed attention entries. Sparse Transformer \citep{SPARSE3} combines strided and local patterns to handle long sequences. Longformer \citep{SPARSE2} introduces a combination of sliding window attention and task-specific global tokens. Subsequently, BigBird \citep{SPARSE5} extends this with random attention connections alongside local and global patterns. Swin \citep{Swin} partitions the input into non-overlapping windows and performs attention within each window. StreamingLLM \citep{StreamingLLM} retains initial attention-sink tokens alongside a sliding window of recent tokens to support bounded-cache streaming inference. These methods achieve subquadratic complexity but rely on fixed sparse patterns that cannot adapt to input-dependent attention distributions, often missing important long-range dependencies. Therefore, CSWin \citep{CSWin} further improves Swin by using different sparse attention in different heads. DuSA \citep{DuSA} introduces a dual-stage sparse attention design to reduce global information loss.

\begin{figure}[htbp]
  \centering
  \includegraphics[width=\linewidth]{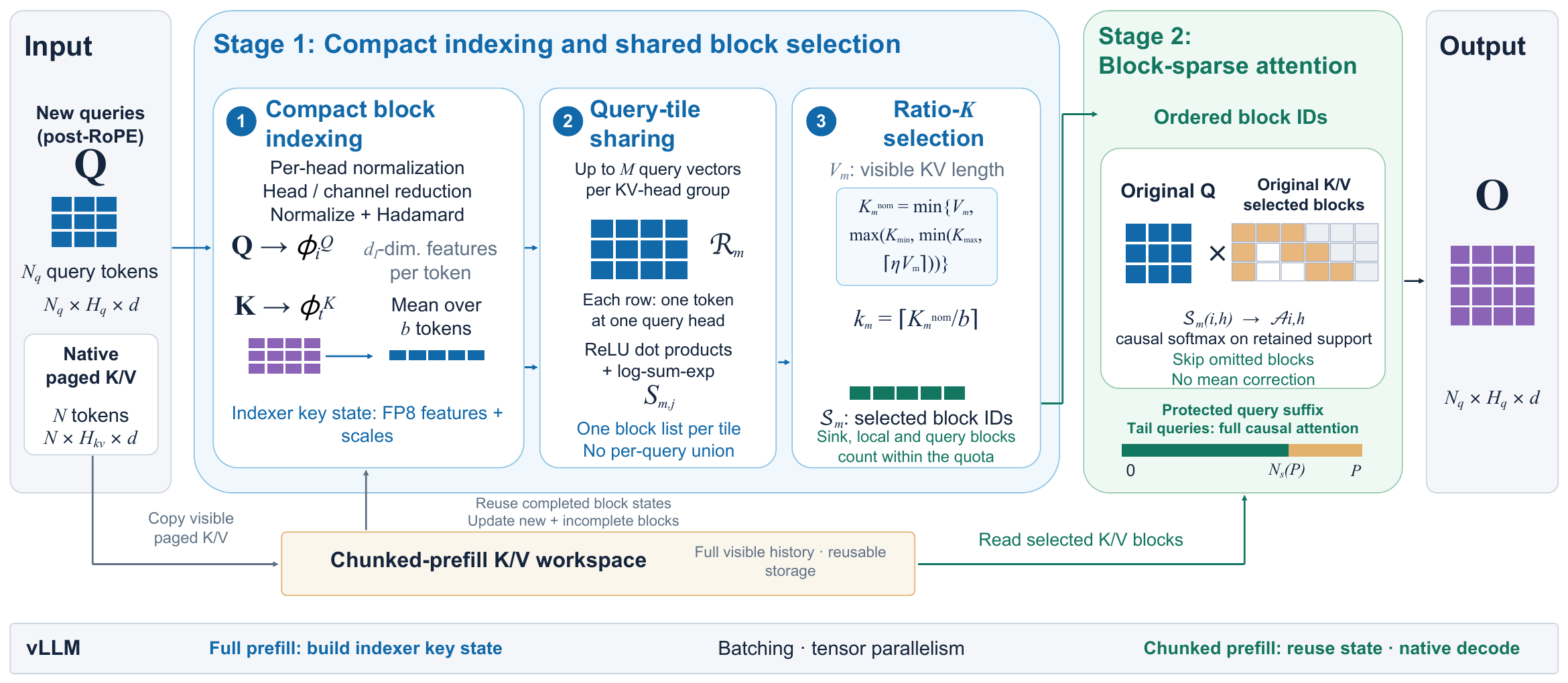}
  \caption{Overview of BFLA. \textbf{Stage 1}: Head-aggregated token features, block pooling, and query-tile sharing produce a shared block set $\mathcal S_m$ using length-adaptive Ratio-$K$ selection, with mandatory blocks within the quota. \textbf{Stage 2}: Block-sparse execution evaluates causal attention over $\mathcal A_{i,h}$, which is derived from the selected block IDs $\mathcal S_{m(i,h)}$, using original Q/K/V. Protected tail queries retain full causal attention instead of Ratio-$K$ selection. The bottom panel summarizes native KV integration and layer--request indexer-state management. Decoding remains native.}
  \label{fig:overview}
\end{figure}

\subsubsection{Dynamic sparse attention}
Dynamic sparse attention usually reduces computation by selecting input-dependent token or block supports. XAttention \citep{xu2025xattention} estimates block importance through antidiagonal scoring and skips low-importance blocks. SpargeAttn \citep{spargeattn} combines sparse-pattern prediction with online softmax-aware filtering to skip unnecessary attention computations. MInference \citep{MInference} assigns A-shape, vertical-slash, or block-sparse patterns to attention heads offline and constructs their sparse indices dynamically during prefilling. FlashPrefill \citep{FlashPrefill} identifies salient attention blocks using mean-pooled keys and query-based scoring, with max-based thresholding replacing Top-$k$ or cumulative-mass selection. FlashPrefill V2 \citep{FlashPrefillV2} extends this approach with pooled-K/V mean correction for discarded blocks, FlashAttention-3 (FA3) aligned sparse kernels, and support for paged KV caches and continuous batching. Quest \citep{Quest} accelerates decoding by selecting KV-cache pages using query-dependent per-page key importance. DeepSeek Sparse Attention (DSA) \citep{DeepSeekV32} uses a learned Lightning Indexer to score preceding tokens and select query-dependent Top-$k$ supports for sparse attention. Qwen Sparse Attention (QSA) \citep{qwen38flashnext} builds on this learned-indexing paradigm by average-pooling projected index keys into micro-blocks, reducing scoring and selection overhead. 

BFLA follows the dynamic sparse attention paradigm and focuses on this computational-efficiency challenge. It co-designs a deterministic Lightning-style block indexer, query-tile-based support sharing, and length-adaptive Ratio-$K$ selection with block-sparse execution, performing attention with the original Q/K/V over the retained support without residual mean correction or model retraining.

\subsection{Linear attention}
Linear attention methods factorize query--key interactions through feature maps so that sequence-length complexity is linear for fixed feature dimensions. The work \cite{TRNN} formulates linear attention with a recurrent representation. Performer approximates the softmax kernel with random features \citep{PEF}. cosFormer uses the cosine-based reweighting feature map \citep{COS}. FLatten develops focused linear attention for vision transformers \citep{flatten}. The work \cite{2SOFT} performs a softmax based feature map on queries and keys, respectively. Their accuracy--efficiency trade-offs depend on the feature map and workload \citep{vanillabest}. Related low-rank approaches include Nystr\"omformer \citep{NTRANS}, which uses a Nystr\"om approximation, CUR-based self-attention approximation \citep{CURSA}, and Primal-Attention \citep{SVDATT}, which uses the primal representation of asymmetric kernel SVD. Mamba instead uses selective state-space modeling, rather than a softmax-attention approximation \citep{Mamba}.

\subsection{Hybrid attention}
Hybrid approaches combine different attention mechanisms or sequence-modeling methods. NSA \citep{NSA} combines compressed attention, top-$k$ selective attention, and sliding window attention in a hardware-aligned hierarchical sparse design. ELFATT \citep{elfatt} uses parallel heads of linear attention and block-sparse attention to capture global and local information. These methods typically require training from scratch or architectural modifications, making them less suitable for plug-and-play inference acceleration of existing LLMs.

\section{Methods}
\label{sec:method}
As illustrated in Fig.~\ref{fig:overview}, BFLA is a training-free two-stage sparse prefill method. Stage~1 constructs compact query features and a block-level indexer key state, then combines query-tile sharing with length-adaptive Ratio-$K$ selection to produce selected block identifiers (IDs). Stage~2 performs causal attention on the selected blocks using the original Q/K/V. The indexer key state supports block scoring. The selected block IDs specify which original K/V blocks are used in attention.

\subsection{Preliminaries}
\label{sec:bfla_preliminaries}
For one request and layer, let $N_q$ be the number of newly scheduled query tokens, $N$ the visible KV length, and $P\ge N$ the full prompt length, with $1\le N_q\le N\le P$. Here $H_q$ and $H_{\mathrm{kv}}$ denote the numbers of query and KV heads, respectively. We use token--head--channel ordering: $\mathbf Q\in\mathbb R^{N_q\times H_q\times d}$ and $\mathbf K,\mathbf V\in\mathbb R^{N\times H_{\mathrm{kv}}\times d}$, where $d$ is the head dimension. All vectors are row vectors. The absolute positions of the newly scheduled query tokens form the set
\[
 \mathcal I=\{N-N_q,\ldots,N-1\},\qquad |\mathcal I|=N_q.
\]
For $i\in\mathcal I$, $0\le h<H_q$, $0\le g<H_{\mathrm{kv}}$, and $0\le t<N$, define
\[
 \mathbf q_{i,h}:=\mathbf Q[i-(N-N_q),h,:],\qquad
 \mathbf k_{t,g}:=\mathbf K[t,g,:],\qquad
 \mathbf v_{t,g}:=\mathbf V[t,g,:].
\]
The first subscript is the absolute token position and the second is an attention-head index. The global logical head counts satisfy $G=H_q/H_{\mathrm{kv}}\in\mathbb N^{+}$, and $\kappa(h)=\lfloor h/G\rfloor$ maps a query head to its KV head. With the positive attention scale $\sigma$, the dense reference is
\begin{equation}
 \mathbf o^{\mathrm{dense}}_{i,h}
 =\sum_{t=0}^{i}
 \frac{\exp(\sigma\mathbf q_{i,h}\mathbf k_{t,\kappa(h)}^\top)}
 {\sum_{v=0}^{i}\exp(\sigma\mathbf q_{i,h}\mathbf k_{v,\kappa(h)}^\top)}
 \mathbf v_{t,\kappa(h)}.
 \label{eq:vsa}
\end{equation}
Fresh full prefill has $N_q=N=P$. BFLA changes the attention execution in targeted attention prefill layers while model weights, decoding, and other layer types remain unchanged.

\subsection{Training-Free Block Indexing}
\label{sec:stage1}
\label{sec:bfla_indexer}
Following the compact-scoring principle of Lightning Indexer \citep{DeepSeekV32}, BFLA replaces learned indexing projections with a deterministic feature map of dimension $d_I$. For a row vector $\mathbf x\in\mathbb R^{d_x}$ with $d_x\ge1$ components, define
\begin{equation}
 \mathcal N_\epsilon(\mathbf x)
 =\frac{\mathbf x}{\sqrt{\|\mathbf x\|_2^2/d_x+\epsilon}},
 \qquad \epsilon>0.
 \label{eq:bfla_rms}
\end{equation}
Here $\epsilon$ is a positive numerical stabilizer. Each head is normalized before averaging, giving one query or key summary per token:
\begin{equation}
 \overline{\mathbf q}_i=\frac1{H_q}
       \sum_{h=0}^{H_q-1}\mathcal N_\epsilon(\mathbf q_{i,h}),
 \qquad
 \overline{\mathbf k}_t=\frac1{H_{\mathrm{kv}}}
       \sum_{g=0}^{H_{\mathrm{kv}}-1}\mathcal N_\epsilon(\mathbf k_{t,g}).
 \label{eq:bfla_head_reduction}
\end{equation}
For this construction, $d_I$ is a power of two dividing $d$. For $\mathbf x\in\mathbb R^d$, let $c=d/d_I$ and reduce adjacent channels by
\begin{equation}
 [\mathcal P(\mathbf x)]_a
 =\frac1{\sqrt c}\sum_{v=ac}^{(a+1)c-1}x_v,
 \qquad 0\le a<d_I.
 \label{eq:bfla_channel_map}
\end{equation}
A second normalization and an orthonormal Walsh--Hadamard matrix $\mathbf H_I\in\mathbb R^{d_I\times d_I}$ produce the indexer features
\begin{equation}
 \boldsymbol\phi_i^Q=\mathcal N_\epsilon(\mathcal P(\overline{\mathbf q}_i))\mathbf H_I,
 \qquad
 \boldsymbol\phi_t^K=\mathcal N_\epsilon(\mathcal P(\overline{\mathbf k}_t))\mathbf H_I.
 \label{eq:bfla_features}
\end{equation}
Partition the visible keys into blocks $\mathcal B_j=\{jb,\ldots,\min((j+1)b,N)-1\}$, where $b$ is the block size, $J=\lceil N/b\rceil$ is the number of key blocks, and $0\le j<J$. Each block is represented by its mean indexer key feature and quantized with a positive per-block scale $a_j$:
\begin{equation}
 \boldsymbol\mu_j
 =\frac1{|\mathcal B_j|}\sum_{t\in\mathcal B_j}\boldsymbol\phi_t^K,
 \qquad
 \widehat{\boldsymbol\mu}_j
 =\mathcal Q_{\mathrm{FP8}}(\boldsymbol\mu_j/a_j).
 \label{eq:bfla_block_index}
\end{equation}
Here $\mathcal Q_{\mathrm{FP8}}$ is elementwise FP8 conversion, with $a_j$ set from the maximum absolute feature value to fit the numerical range. The reconstructed feature $a_j\widehat{\boldsymbol\mu}_j\approx\boldsymbol\mu_j$ is used only for ranking. The indexer key state stores each block's quantized key feature $\widehat{\boldsymbol\mu}_j$ and scale $a_j$ for relevance scoring. Pooling reduces the token axis of this state. 

\subsection{Query-Tile Sharing and \texorpdfstring{Ratio-$K$}{Ratio-K} Selection}
\label{sec:bfla_query_sharing}
\label{sec:bfla_ratio_k}

\paragraph{Sharing a block selection.}
BFLA selects one KV-block list for several query vectors rather than selecting tokens independently and merging their lists afterward. For each KV head $g$, its $G$ query heads are $gG,\ldots,gG+G-1$. Their query vectors are ordered first by token position, then by head: each row is one $d$-dimensional query vector for one token at one head. Let $M$ be the maximum number of query vectors in a logical tile. Consecutive groups of at most $M$ rows form these tiles. Larger tiles amortize selection and metadata costs over more rows, but require more query vectors to share a selection. Thus, $M$ controls a quality--efficiency trade-off, independently of the indexer feature dimension $d_I$; it need not equal the physical compute-tile size.

For a contiguous interval of current queries that require block selection, let $i_0$ be its first absolute token position, and $n_{\mathrm{seg}}$ be its number of query tokens. Thus, $\{i_0,\ldots,i_0+n_{\mathrm{seg}}-1\}\subseteq\mathcal I$. For token $i$ and within-group head $s$, the row number and its token position are $r=(i-i_0)G+s$ and $i(r)=i_0+\lfloor r/G\rfloor$, where $h=gG+s$ and $s=h\bmod G\in\{0,\ldots,G-1\}$. All token and head indices are integers. For $0\le m<\lceil n_{\mathrm{seg}}G/M\rceil$, the valid row numbers of tile $m$ are
\begin{equation}
 \mathcal R_m=\{r:mM\le r<\min((m+1)M,n_{\mathrm{seg}}G)\}.
 \label{eq:bfla_query_tile}
\end{equation}
When $G$ divides $M$, a full tile contains $M/G$ token positions. Otherwise, a token's heads may lie in adjacent tiles and use their respective selected lists. The final tile may contain fewer than $M$ valid rows.

\paragraph{Shared block ranking.}
Let $f_m$ and $e_m$ be the first and last absolute token positions represented in $\mathcal R_m$. The candidate blocks are $\mathcal C_m=\{0,\ldots,\lfloor e_m/b\rfloor\}$. Using $[x]_+=\max(x,0)$, compute
\begin{equation}
 \begin{aligned}
 u_{i,j}&=\frac{a_j}{\sqrt{d_I}}
          [\boldsymbol\phi_i^Q\widehat{\boldsymbol\mu}_j^\top]_+,\\
 S_{m,j}&=\log\sum_{r\in\mathcal R_m}\exp(u_{i(r),j}).
 \end{aligned}
 \label{eq:bfla_shared_score}
\end{equation}
Each query token has one head-aggregated indexer feature, so it contributes once for each of its head rows in the tile. The log-sum-exp combines these contributions into one block ranking, reused across KV-head groups.

\paragraph{Length-adaptive Ratio-$K$ selection.}
For visible length $V_m=e_m+1$ of tile $m$, Ratio-$K$ selection sets a nominal token budget $K_m^{\mathrm{nom}}$ and converts it into a block count $k_m$:
\begin{equation}
 \begin{aligned}
 K_m^{\mathrm{nom}}
 &=\min\{V_m,\max(K_{\min},\min(K_{\max},\lceil\eta V_m\rceil))\},\\
 k_m&=\left\lceil K_m^{\mathrm{nom}}/b\right\rceil.
 \end{aligned}
 \label{eq:bfla_ratio_budget}
\end{equation}
Here $\eta\in(0,1]$ is the context-length ratio, and the positive integers $K_{\min}$ and $K_{\max}$ are the lower and upper token-budget limits before clipping to $V_m$, with $K_{\min}\le K_{\max}$. For the variant without an upper selection limit, the inner $\min(K_{\max},\cdot)$ in Eq.~\eqref{eq:bfla_ratio_budget} is omitted. The budget adapts to the visible context length rather than the cumulative attention mass.

Let $\mathcal M_m\subseteq\mathcal C_m$ contain the mandatory blocks covering initial sink tokens, a local band, and the tile's query range. They count toward the same quota, and the remaining slots are filled by the highest-scoring nonmandatory blocks:
\begin{equation}
 \mathcal S_m=\mathcal M_m\cup
 \operatorname{Top}_{k_m-|\mathcal M_m|}
 \{(j,S_{m,j}):j\in\mathcal C_m\setminus\mathcal M_m\}.
 \label{eq:bfla_topk}
\end{equation}
Only feasible configurations with $|\mathcal M_m|\le k_m\le|\mathcal C_m|$ are considered, yielding $|\mathcal S_m|=k_m$. The selector returns integer block IDs forming $\mathcal S_m$. We use $\operatorname{Top}_0=\varnothing$. Overlaps among mandatory rules count once. Varying tile budgets are handled together by batched Top-$k$ selection. Whole-block rounding adds at most $b-1$ token slots before causal masking.

\paragraph{Full-support protection.}
When $N\le K_{\min}$, the Ratio-$K$ budget covers all causal candidates, so all current queries use full causal attention without score-based selection. For longer contexts, queries at positions $N_s(P),\ldots,P-1$ use full causal attention, while the remaining query tiles follow Ratio-$K$ selection. The boundary is defined in Appendix~\ref{app:bfla_policy_details} using the complete prompt and is not reset at prefill chunk boundaries. Setting $r_{\mathrm{tail}}=0$ disables only suffix protection.

\subsection{Support-Restricted Block-Sparse Attention}
\label{sec:stage2}
\label{sec:bfla_execution}
For queries requiring selection, $m(i,h)$ denotes the logical tile containing the query vector $\mathbf q_{i,h}$, where $i$ is the absolute token position and $h$ is the query-head index. The retained KV-token positions are
\begin{equation}
 \mathcal A_{i,h}=
 \begin{cases}
  \{0,\ldots,i\},
  &N\le K_{\min}\ \text{or}\ i\ge N_s(P),\\
  \{t:0\le t\le i,\ \lfloor t/b\rfloor\in\mathcal S_{m(i,h)}\},
  &\text{otherwise}.
 \end{cases}
 \label{eq:bfla_support}
\end{equation}
Thus, protected queries use full causal attention, while other queries attend to the causal positions in their selected blocks. BFLA computes
\begin{equation}
 \mathbf o_{i,h}
 =\sum_{t\in\mathcal A_{i,h}}
 \frac{\exp(\sigma\mathbf q_{i,h}\mathbf k_{t,\kappa(h)}^\top)}
 {\sum_{v\in\mathcal A_{i,h}}
      \exp(\sigma\mathbf q_{i,h}\mathbf k_{v,\kappa(h)}^\top)}
 \mathbf v_{t,\kappa(h)}.
 \label{eq:bfla_output}
\end{equation}
Each query head uses the original Q/K/V and normalizes attention over its retained positions. Omitted blocks contribute neither values nor mean correction.

Building on the FlashAttention-3-derived sparse pipeline of FlashPrefill V2 \citep{Flash3,FlashPrefillV2}, we implement BFLA's block-sparse execution driven by the selected block IDs. The attention loop processes only selected blocks and reuses their K/V across query vectors within a compute tile. Physical compute tiles may subdivide a logical query tile without changing its selected blocks. The same execution framework handles sparse queries and protected full-support queries.

\subsection{vLLM Integration and Indexer State}
\label{sec:bfla_runtime}
Our vLLM integration preserves native KV-cache management and decoding. Full prefill constructs the indexer key state from the current keys. For chunked prefill, the state is maintained per layer and request: completed blocks are reused, while new and incomplete blocks are updated. Each query tile requiring selection uses the available state to compute block scores and select its shared block list. When attention heads are distributed across GPUs, compact feature contributions are summed before the second normalization and Hadamard transform, counting replicated KV heads once. The chunked-prefill adapter copies the visible paged K/V into reusable contiguous storage for sparse execution.

\subsection{Computation and Memory}
\label{sec:bfla_complexity}
For one request and layer, a prefill call processes $N_q$ new query tokens against a visible KV history of length $N$. Their absolute token positions are $i=N-N_q,\ldots,N-1$. The computational complexity of dense attention (GQA) is $O(H_qN_qNd)$.

Let $T_s$ be the number of logical query tiles requiring block selection in the current prefill step. Each tile contains at most $M$ token--head query rows and scores at most $J=\lceil N/b\rceil$ key blocks using $d_I$-dimensional indexer features. Block scoring therefore requires at most
\begin{equation}
 O(T_s M J d_I)
 \label{eq:bfla_scoring_cost}
\end{equation}
operations. The head-aggregated features produce one shared score vector per tile, rather than a separate vector for each KV head. Sharing reduces the number of independent selections and their metadata.

For query position $i$ and query head $h$, $|\mathcal A_{i,h}|$ is the number of retained causal KV-token positions defined by Eq.~\eqref{eq:bfla_support}. The logical attention work is therefore
\begin{equation}
 O\!\left(
 d\sum_{i=N-N_q}^{N-1}\sum_{h=0}^{H_q-1}|\mathcal A_{i,h}|
 \right).
 \label{eq:bfla_execution_cost}
\end{equation}
This sum includes protected full-support queries, not only score-selected queries. For full prefill, $N_q=N$, so the query index runs from $0$ to $N-1$; block scoring remains quadratic in $N$ for fixed tiling parameters, even after the selection budget saturates. For each layer--request pair, the indexer key state stores $d_I$ FP8 values and one scale per block, requiring $O(Jd_I)$ storage. 

\section{Balancing Indexing Cost and Attention Coverage}
\label{sec:theory}
In this section, we analyze how indexer accuracy and the retention budget jointly affect attention output error. For the budgets chosen by Ratio-$K$ selection, our bound states a sufficient condition under which retaining more blocks compensates for less accurate indexer scores. The analysis concerns attention outputs before model-specific output gating and projection. We use the notation of Section~\ref{sec:method} and compare dense and sparse attention in real arithmetic on the same fixed, finite Q/K/V for one prefill call, with query positions $i\in\mathcal I$.

\paragraph{Budget and score distortion.}
Fix a query/head pair $(i,h)$ in a score-selected tile $m=m(i,h)$. Let $\delta_{i,h}$ be the sum of original dense attention weights omitted by its support, and let $D^V_{i,h}$ be the largest Euclidean distance between its causally available value vectors. Renormalizing the retained weights gives
\[
 \|\mathbf o^{\mathrm{dense}}_{i,h}-\mathbf o_{i,h}\|_2
 \le D^V_{i,h}\delta_{i,h}
\]
(Lemma~\ref{lem:bfla_mass_identity}). We therefore analyze how ranking and budget control $\delta_{i,h}$.

Let $\mathcal F_m=\mathcal C_m\setminus\mathcal M_m$ be the set of nonmandatory candidates, containing $J_m^{\mathrm{free}}=|\mathcal F_m|$ blocks. After accounting for mandatory blocks, the available quota is $k_m^{\mathrm{free}}=k_m-|\mathcal M_m|$. The query-range protection ensures that every nonmandatory block is fully causal for every row in the tile. Mandatory blocks preserve a fraction $\beta_{i,h}$ of the original dense attention mass. Among the nonmandatory blocks, the best possible selection keeps the $k$ blocks with the largest attention mass. We denote the fraction of nonmandatory attention mass left out by this ideal selection as $\tau_{i,h}(k)$. This oracle tail measures the unavoidable loss of attention mass from retaining only $k$ of these blocks.

For query $i$, head $h$, and a nonmandatory block $j\in\mathcal F_m$, let $\zeta_{i,h,j}$ be the logarithm of its total unnormalized attention weight. Higher scores correspond to blocks with more dense attention mass. For nonempty $\mathcal F_m$ and finite shared scores $S_{m,j}$, compare these reference scores up to a common additive shift, which leaves Top-$k$ unchanged.
\begin{equation}
 \varepsilon^{\mathrm{rank}}_{i,h}
 =\inf_{a_{\mathrm{shift}}\in\mathbb R}
   \max_{j\in\mathcal F_m}
   |\zeta_{i,h,j}-S_{m,j}-a_{\mathrm{shift}}|.
 \label{eq:bfla_main_distortion}
\end{equation}
This is a discrepancy in score values, not a count of ranking errors; it can be nonzero even when the ordering is correct. It accounts for compact indexing and differences among queries and heads sharing the scores. 

\begin{bflaboundtheorem}[Indexer--budget error bound]
\label{thm:bfla_indexer_budget}
For finite indexer scores and $1\le k_m^{\mathrm{free}}<J_m^{\mathrm{free}}$, BFLA's Top-$k$ selection on $\mathcal F_m$ satisfies
\begin{equation}
 \begin{aligned}
 \delta_{i,h}
 &\le (1-\beta_{i,h})
       \min\!\left\{1,
       e^{2\varepsilon^{\mathrm{rank}}_{i,h}}
      \tau_{i,h}(k_m^{\mathrm{free}})\right\},\\
 \|\mathbf o^{\mathrm{dense}}_{i,h}-\mathbf o_{i,h}\|_2
 &\le D^V_{i,h}(1-\beta_{i,h})
       \min\!\left\{1, 
       e^{2\varepsilon^{\mathrm{rank}}_{i,h}}
       \tau_{i,h}(k_m^{\mathrm{free}})\right\}.
 \end{aligned}
 \label{eq:bfla_main_tradeoff}
\end{equation}
\end{bflaboundtheorem}

The result combines normalized-support truncation \citep{TopKTV,Twilight} with a comparison between surrogate and oracle selection. Lemma~\ref{lem:bfla_mass_identity} and Theorem~\ref{thm:bfla_ratio_error}, together with $\delta_{i,h}\le1-\beta_{i,h}$, prove the bound; empty and full free-block selections are handled in Appendix~\ref{app:bfla_budget_oracle}.

\paragraph{Balancing the indexer and retention budget.}
Consider two configurations with the same Q/K/V, candidate blocks, and mandatory blocks. Let their score distortions be $\varepsilon_1,\varepsilon_2$ and their free budgets be $1\le k_1\le k_2<J_m^{\mathrm{free}}$. A sufficient condition for the second configuration to have no larger tail-based bound (and hence no larger clipped bound) is
\begin{equation} \tau_{i,h}(k_2)\le e^{-2(\varepsilon_2-\varepsilon_1)}\tau_{i,h}(k_1). \label{eq:bfla_main_compensation} \end{equation} 
Thus, a cheaper indexer can tolerate greater score distortion when the larger budget reduces the oracle tail sufficiently. Under its score-gap condition, Corollary~\ref{cor:bfla_oversampling} guarantees that the indexer retains the $s$ free blocks with the largest true attention mass, even if it ranks them in the wrong order. If the condition holds for every query/head row sharing a tile, their shared selection retains each row's top-$s$ blocks. The bound motivates joint selection of the indexer and retention budget. Section~\ref{sec:experiments} evaluates the resulting accuracy and latency.

\section{Experiments}
\label{sec:experiments}

\subsection{Experimental Setup}
\label{sec:bfla_experimental_setup}

\paragraph{Models and baselines.}
We evaluate Gemma4-E4B-IT \citep{gemma4}, Qwen3.6-27B \citep{qwen36}, Qwen3.8-27B\footnote{\url{https://huggingface.co/Qwen/Qwen3.8-27B}}, and Qwen3.8-Flash-Next \citep{qwen38flashnext} on LongBench-v2 \citep{LongBenchV2}, RULER \citep{RULER}, and InfiniteBench \citep{InfiniteBench}, and additionally evaluate Qwen3.8-Flash-Next on Terminal Bench 2.1 \citep{merrill2026terminalbench}. Experiments use four H200 GPUs and unchanged model weights. BFLA replaces only the targeted prefill attention while decoding, local or recurrent layers, and native KV-sharing semantics remain unchanged. We compare against dense attention and FlashPrefill V2 \citep{FlashPrefillV2} on Gemma and the two Qwen-27B models, and against dense attention and QSA on Qwen3.8-Flash-Next. The dense baseline uses vLLM-native Triton attention for Gemma and FlashAttention-3 for Qwen.

\paragraph{Shared sparsity configuration.}
BFLA uses one configuration without per-model retuning: an indexer feature dimension of $d_I=128$ with FP8 key-state storage, $b=64$ tokens per key block, $M=128$ token--head query vectors per logical tile, $K_{\min}=2048$, $K_{\max}=32768$, and Ratio-$K$ with $\eta=1/3$. Original attention Q/K/V remain BF16. Mandatory sink and local blocks count toward the selection budget. Tail protection covers the final $r_{\mathrm{tail}}=8$ reference groups of $M_{\mathrm{tail}}=128$ query vectors per KV-head group. Sharing is within query tiles, not across layers. Appendix~\ref{sec:bfla_ablation} examines the indexer feature dimension and storage precision, block size, sharing size, selection budgets, and tail-protection count. We follow the upstream reference configuration of FlashPrefill V2: $\alpha=0.1$, 128-token selection blocks, two sink blocks, a four-block local window, eight full-support trailing PackGQA query tiles, and enabled mean correction, without model-specific hyperparameter tuning. 

\paragraph{Long-context evaluation protocol.}
All three benchmarks use greedy generation with thinking disabled, seed~42, normal EOS termination, and one active request per model replica. Inputs and output limits are matched across methods within each model. LongBench-v2 covers all 503 questions with the official non-CoT template and a maximum output length of 2,048 tokens. RULER covers 13 tasks at 32K, 64K, and 128K with 500 examples per task and length, without input truncation. InfiniteBench covers all 12 tasks and 3,946 examples using the official task templates. RULER and InfiniteBench retain task-specific output limits and supplied examples. Gemma and the two Qwen-27B models use BF16 weights on one GPU with a 128K context limit and full-sequence prefill. Qwen3.8-Flash-Next uses FP8 weights on two GPUs with tensor parallelism, a 256K context limit, and a 16K-token scheduling budget for chunked prefill. Overlong LongBench-v2 and InfiniteBench inputs are head--tail truncated after reserving output space. Here, 1K denotes 1,024 tokens.

\paragraph{Metrics.}
LongBench-v2 uses accuracy, with unparseable answers counted as incorrect; its scores and latencies are averaged over three complete runs. RULER averages scores over tasks and lengths, and InfiniteBench reports the equal-weight mean of its 12 task scores. We report mean time to first token (TTFT) and end-to-end request latency (E2E), with speedups computed from summed latencies over matched examples. Operator measurements cover indexing, selection, and attention.

\subsection{LongBench-v2}
\label{sec:bfla_quality}

\begin{table*}[htbp]
\centering
\small
\caption{LongBench-v2 accuracy, averaged over three runs with a 2K-token maximum output length.}
\label{tab:bfla_longbench_quality}
\begin{tabular}{llr}
\toprule
Model & Method & Accuracy (\%) \\
\midrule
Gemma4-E4B-IT
 & Dense (Triton) & 35.65 \\
 & FlashPrefill V2 & 34.66 \\
 & BFLA (ours) & 37.11 \\
\midrule
Qwen3.6-27B
 & Dense (FA3) & 54.01 \\
 & FlashPrefill V2 & 51.29 \\
 & BFLA (ours) & 53.48 \\
\midrule
Qwen3.8-27B
 & Dense (FA3) & 52.42 \\
 & FlashPrefill V2 & 46.45 \\
 & BFLA (ours) & 51.69 \\
\midrule
Qwen3.8-Flash-Next
 & Dense (FA3) & 57.32 \\
 & QSA & 56.99 \\
 & BFLA (ours) & 57.32 \\
\bottomrule
\end{tabular}
\end{table*}

BFLA changes accuracy relative to the dense baselines by $+1.46$, $-0.53$, $-0.73$, and $0.00$ percentage points on Gemma, Qwen3.6-27B, Qwen3.8-27B, and Qwen3.8-Flash-Next, respectively (Table~\ref{tab:bfla_longbench_quality}). It exceeds FlashPrefill V2 by 2.19--5.24 points on the first three models and QSA by 0.33 points on Qwen3.8-Flash-Next. BFLA achieves TTFT speedups of $1.224$--$3.846\times$ and E2E speedups of $1.059$--$1.738\times$ over the dense baselines (Table~\ref{tab:bfla_longbench_runtime}). FlashPrefill V2 has lower TTFT on Gemma and the two Qwen-27B models, whereas BFLA has higher accuracy and lower E2E on all three. On Qwen3.8-Flash-Next, BFLA gains higher accuracy over QSA with slightly higher TTFT and E2E.

\begin{table*}[htbp]
\centering
\small
\setlength{\tabcolsep}{4pt}
\caption{LongBench-v2 mean request latencies with a 2K-token maximum output length, averaged over all 503 questions and three runs. Speedups are ratios of summed latencies relative to dense baselines. TTFT is the primary metric for prefill efficiency. E2E additionally reflects response length and decoding. Qwen3.8-Flash-Next retains QSA decoding throughout.}
\label{tab:bfla_longbench_runtime}
\begin{tabular}{llrrrr}
\toprule
 & & \multicolumn{2}{c}{Mean latency (s)}
 & \multicolumn{2}{c}{Speedup over Dense} \\
\cmidrule(lr){3-4}\cmidrule(lr){5-6}
Model & Method & TTFT & E2E & TTFT & E2E \\
\midrule
Gemma4-E4B-IT
 & Dense (Triton)
 & 8.891 & 15.337 & $1.000\times$ & $1.000\times$ \\
 & FlashPrefill V2
 & 2.218 & 8.928 & $4.009\times$ & $1.718\times$ \\
 & BFLA (ours)
 & 2.312 & 8.824 & $3.846\times$ & $1.738\times$ \\
\midrule
Qwen3.6-27B
 & Dense (FA3)
 & 10.043 & 16.447 & $1.000\times$ & $1.000\times$ \\
 & FlashPrefill V2
 & 7.465 & 15.454 & $1.345\times$ & $1.064\times$ \\
 & BFLA (ours)
 & 8.204 & 14.902 & $1.224\times$ & $1.104\times$ \\
\midrule
Qwen3.8-27B
 & Dense (FA3)
 & 10.069 & 21.999 & $1.000\times$ & $1.000\times$ \\
 & FlashPrefill V2
 & 7.482 & 21.451 & $1.346\times$ & $1.026\times$ \\
 & BFLA (ours)
 & 8.206 & 20.767 & $1.227\times$ & $1.059\times$ \\
\midrule
Qwen3.8-Flash-Next
 & Dense (FA3)
 & 6.463 & 9.264 & $1.000\times$ & $1.000\times$ \\
 & QSA
 & 4.338 & 7.333 & $1.490\times$ & $1.263\times$ \\
 & BFLA (ours)
 & 4.504 & 7.391 & $1.435\times$ & $1.253\times$ \\
\bottomrule
\end{tabular}
\end{table*}

\subsection{RULER}
\label{sec:bfla_ruler_quality}
\label{sec:bfla_ruler_runtime}

\begin{table*}[htbp]
\centering
\small
\setlength{\tabcolsep}{4pt}
\caption{RULER scores (0--100), with 500 examples per task and length. Overall is the equal-weight mean of 39 task--length combinations. The alternative is FlashPrefill V2 for the first three models and QSA for Qwen3.8-Flash-Next.}
\label{tab:bfla_ruler_quality}
\begin{tabular}{llrrr}
\toprule
Model & Length & \shortstack{Dense\\(Triton/FA3)} & FPv2/QSA & BFLA (ours) \\
\midrule
Gemma4-E4B-IT & 32K & 94.61 & 91.24 & 94.02 \\
 & 64K & 91.88 & 88.28 & 92.51 \\
 & 128K & 86.16 & 81.48 & 88.84 \\
 & Overall & 90.89 & 87.00 & 91.79 \\
\midrule
Qwen3.6-27B & 32K & 97.17 & 95.67 & 97.04 \\
 & 64K & 97.02 & 94.40 & 96.79 \\
 & 128K & 96.71 & 89.24 & 96.45 \\
 & Overall & 96.96 & 93.10 & 96.76 \\
\midrule
Qwen3.8-27B & 32K & 96.81 & 95.06 & 96.58 \\
 & 64K & 96.60 & 93.21 & 96.27 \\
 & 128K & 96.19 & 89.62 & 96.01 \\
 & Overall & 96.53 & 92.63 & 96.28 \\
\midrule
Qwen3.8-Flash-Next & 32K & 93.80 & 92.37 & 93.09 \\
 & 64K & 95.55 & 92.13 & 95.30 \\
 & 128K & 94.59 & 92.67 & 94.68 \\
 & Overall & 94.65 & 92.39 & 94.36 \\
\bottomrule
\end{tabular}
\end{table*}

\begin{table*}[htbp]
\centering
\small
\setlength{\tabcolsep}{3pt}
\caption{RULER task scores (0--100), averaged equally over 32K, 64K, and 128K. NIAH, S, MK, MV, and MQ denote needle-in-a-haystack, single-needle, multi-key, multi-value, and multi-query retrieval; VT, CWE, FWE, and QA denote variable tracking, common-word extraction, frequent-word extraction, and question answering.}
\label{tab:bfla_ruler_tasks}
\begin{tabular}{l*{12}{r}}
\toprule
 & \multicolumn{3}{c}{Gemma4-E4B-IT}
 & \multicolumn{3}{c}{Qwen3.6-27B}
 & \multicolumn{3}{c}{Qwen3.8-27B}
 & \multicolumn{3}{c}{Qwen3.8-Flash-Next} \\
\cmidrule(lr){2-4}\cmidrule(lr){5-7}\cmidrule(lr){8-10}\cmidrule(lr){11-13}
Task & \shortstack{Dense\\(Triton)} & FPv2 & BFLA & \shortstack{Dense\\(FA3)} & FPv2 & BFLA & \shortstack{Dense\\(FA3)} & FPv2 & BFLA & \shortstack{Dense\\(FA3)} & QSA & BFLA \\
\midrule
NIAH-S1 & 100.00 & 100.00 & 100.00 & 100.00 & 100.00 & 100.00 & 100.00 & 100.00 & 100.00 & 100.00 & 100.00 & 100.00 \\
NIAH-S2 & 100.00 & 100.00 & 100.00 & 100.00 & 100.00 & 100.00 & 100.00 & 100.00 & 100.00 & 99.93 & 99.93 & 99.93 \\
NIAH-S3 & 100.00 & 100.00 & 100.00 & 100.00 & 99.93 & 100.00 & 100.00 & 100.00 & 100.00 & 100.00 & 100.00 & 100.00 \\
NIAH-MK1 & 99.73 & 99.00 & 99.40 & 100.00 & 99.20 & 100.00 & 100.00 & 99.33 & 99.93 & 100.00 & 99.93 & 100.00 \\
NIAH-MK2 & 99.80 & 94.40 & 99.40 & 100.00 & 99.80 & 100.00 & 99.93 & 99.60 & 100.00 & 99.93 & 100.00 & 99.80 \\
NIAH-MK3 & 98.73 & 97.20 & 98.13 & 100.00 & 100.00 & 100.00 & 100.00 & 99.60 & 100.00 & 99.87 & 100.00 & 98.13 \\
NIAH-MV & 90.35 & 85.35 & 91.72 & 100.00 & 74.77 & 98.65 & 100.00 & 69.15 & 97.92 & 100.00 & 100.00 & 99.47 \\
NIAH-MQ & 99.38 & 99.00 & 99.52 & 100.00 & 99.93 & 100.00 & 100.00 & 99.68 & 100.00 & 82.00 & 60.72 & 79.60 \\
VT & 96.36 & 57.09 & 94.72 & 100.00 & 95.88 & 100.00 & 100.00 & 98.40 & 99.99 & 100.00 & 100.00 & 99.95 \\
CWE & 68.53 & 69.76 & 82.62 & 100.00 & 85.47 & 99.99 & 99.99 & 87.57 & 99.97 & 99.99 & 99.98 & 99.98 \\
FWE & 96.76 & 97.33 & 96.91 & 99.73 & 98.16 & 99.71 & 99.67 & 99.36 & 99.62 & 96.04 & 96.04 & 98.24 \\
QA1 & 79.47 & 80.80 & 80.07 & 89.93 & 90.20 & 89.87 & 86.60 & 86.40 & 86.80 & 86.73 & 83.13 & 86.87 \\
QA2 & 52.40 & 51.07 & 50.80 & 70.87 & 67.00 & 69.67 & 68.73 & 65.13 & 67.47 & 65.93 & 61.33 & 64.67 \\
\midrule
Overall & 90.89 & 87.00 & 91.79 & 96.96 & 93.10 & 96.76 & 96.53 & 92.63 & 96.28 & 94.65 & 92.39 & 94.36 \\
\bottomrule
\end{tabular}
\end{table*}

\begin{table*}[htbp]
\centering
\small
\setlength{\tabcolsep}{4pt}
\caption{RULER: Mean request latencies and speedups over dense baselines at each context length.}
\label{tab:bfla_ruler_runtime}
\begin{tabular}{lllrrrr}
\toprule
 & & & \multicolumn{2}{c}{Mean latency (s)} & \multicolumn{2}{c}{Speedup} \\
\cmidrule(lr){4-5}\cmidrule(lr){6-7}
Model & Length & Method & TTFT & E2E & TTFT & E2E \\
\midrule
Gemma4-E4B-IT & 32K & Dense (Triton) & 1.252 & 1.480 & $1.000\times$ & $1.000\times$ \\
 &  & FlashPrefill V2 & 0.679 & 0.898 & $1.844\times$ & $1.648\times$ \\
 &  & BFLA (ours) & 0.674 & 0.893 & $1.858\times$ & $1.657\times$ \\
\addlinespace[2pt]
 & 64K & Dense (Triton) & 4.076 & 4.366 & $1.000\times$ & $1.000\times$ \\
 &  & FlashPrefill V2 & 1.569 & 1.847 & $2.598\times$ & $2.364\times$ \\
 &  & BFLA (ours) & 1.547 & 1.826 & $2.635\times$ & $2.391\times$ \\
\addlinespace[2pt]
 & 128K & Dense (Triton) & 14.985 & 15.383 & $1.000\times$ & $1.000\times$ \\
 &  & FlashPrefill V2 & 3.964 & 4.349 & $3.780\times$ & $3.537\times$ \\
 &  & BFLA (ours) & 3.751 & 4.137 & $3.995\times$ & $3.718\times$ \\
\midrule
Qwen3.6-27B & 32K & Dense (FA3) & 2.888 & 3.268 & $1.000\times$ & $1.000\times$ \\
 &  & FlashPrefill V2 & 2.703 & 3.082 & $1.068\times$ & $1.060\times$ \\
 &  & BFLA (ours) & 2.775 & 3.163 & $1.041\times$ & $1.033\times$ \\
\addlinespace[2pt]
 & 64K & Dense (FA3) & 6.490 & 6.885 & $1.000\times$ & $1.000\times$ \\
 &  & FlashPrefill V2 & 5.511 & 5.901 & $1.178\times$ & $1.167\times$ \\
 &  & BFLA (ours) & 5.834 & 6.238 & $1.112\times$ & $1.104\times$ \\
\addlinespace[2pt]
 & 128K & Dense (FA3) & 15.632 & 16.056 & $1.000\times$ & $1.000\times$ \\
 &  & FlashPrefill V2 & 11.190 & 11.610 & $1.397\times$ & $1.383\times$ \\
 &  & BFLA (ours) & 12.570 & 13.003 & $1.244\times$ & $1.235\times$ \\
\midrule
Qwen3.8-27B & 32K & Dense (FA3) & 2.887 & 3.277 & $1.000\times$ & $1.000\times$ \\
 &  & FlashPrefill V2 & 2.708 & 3.098 & $1.066\times$ & $1.058\times$ \\
 &  & BFLA (ours) & 2.740 & 3.140 & $1.054\times$ & $1.044\times$ \\
\addlinespace[2pt]
 & 64K & Dense (FA3) & 6.491 & 6.891 & $1.000\times$ & $1.000\times$ \\
 &  & FlashPrefill V2 & 5.530 & 5.921 & $1.174\times$ & $1.164\times$ \\
 &  & BFLA (ours) & 5.783 & 6.192 & $1.122\times$ & $1.113\times$ \\
\addlinespace[2pt]
 & 128K & Dense (FA3) & 15.637 & 16.060 & $1.000\times$ & $1.000\times$ \\
 &  & FlashPrefill V2 & 11.250 & 11.657 & $1.390\times$ & $1.378\times$ \\
 &  & BFLA (ours) & 12.483 & 12.918 & $1.253\times$ & $1.243\times$ \\
\midrule
Qwen3.8-Flash-Next & 32K & Dense (FA3) & 1.074 & 1.199 & $1.000\times$ & $1.000\times$ \\
 &  & QSA & 1.070 & 1.190 & $1.004\times$ & $1.008\times$ \\
 &  & BFLA (ours) & 1.017 & 1.138 & $1.056\times$ & $1.054\times$ \\
\addlinespace[2pt]
 & 64K & Dense (FA3) & 2.425 & 2.548 & $1.000\times$ & $1.000\times$ \\
 &  & QSA & 2.186 & 2.296 & $1.109\times$ & $1.110\times$ \\
 &  & BFLA (ours) & 2.150 & 2.268 & $1.128\times$ & $1.123\times$ \\
\addlinespace[2pt]
 & 128K & Dense (FA3) & 5.891 & 5.987 & $1.000\times$ & $1.000\times$ \\
 &  & QSA & 4.466 & 4.562 & $1.319\times$ & $1.312\times$ \\
 &  & BFLA (ours) & 4.629 & 4.725 & $1.273\times$ & $1.267\times$ \\
\bottomrule
\end{tabular}
\end{table*}

BFLA's overall scores are 91.79, 96.76, 96.28, and 94.36 on the four models, respectively, exceeding the corresponding sparse baseline in every case (Tables~\ref{tab:bfla_ruler_quality} and~\ref{tab:bfla_ruler_tasks}). Gemma's gain over the dense baseline comes mainly from common-word extraction. On Qwen3.8-Flash-Next, BFLA reaches 79.60 on multi-query retrieval, compared with 60.72 for QSA and 82.00 for the dense baseline. At 128K, BFLA achieves TTFT speedups of $1.244$--$3.995\times$ and E2E speedups of $1.235$--$3.718\times$ over the dense baselines (Table~\ref{tab:bfla_ruler_runtime}). It is faster than FlashPrefill V2 on Gemma; FlashPrefill V2 remains faster on the two Qwen-27B models with lower scores. On Qwen3.8-Flash-Next, BFLA is faster than QSA at 32K and 64K but slightly slower at 128K.

\subsection{InfiniteBench}
\label{sec:bfla_infinitebench_quality}
\label{sec:bfla_infinitebench_runtime}
BFLA achieves macro-average scores of 42.25, 57.56, 58.56, and 61.39 on the four models (Table~\ref{tab:bfla_infinitebench_quality}). It exceeds FlashPrefill V2 by 1.26, 0.60, and 3.53 points in Gemma and the two Qwen-27B models, with notable gains in KV retrieval and dialogue. On Qwen3.8-Flash-Next, BFLA scores above the dense baseline's 61.14 but below QSA's 62.89. BFLA achieves TTFT speedups of $1.229$--$3.906\times$ and E2E speedups of $1.133$--$3.222\times$ over the dense baselines (Table~\ref{tab:bfla_infinitebench_runtime}). It also reduces TTFT and E2E relative to FlashPrefill V2 on Gemma. FlashPrefill V2 remains faster on the two Qwen-27B models. QSA has slightly lower TTFT and E2E on Qwen3.8-Flash-Next than those of BFLA.

\begin{table*}[htbp]
\centering
\small
\setlength{\tabcolsep}{2.5pt}
\caption{InfiniteBench task scores (0--100). MC and Calc denote multiple-choice and calculation.}
\label{tab:bfla_infinitebench_quality}
\begin{tabular}{l*{12}{r}}
\toprule
 & \multicolumn{3}{c}{Gemma4-E4B-IT}
 & \multicolumn{3}{c}{Qwen3.6-27B}
 & \multicolumn{3}{c}{Qwen3.8-27B}
 & \multicolumn{3}{c}{Qwen3.8-Flash-Next} \\
\cmidrule(lr){2-4}\cmidrule(lr){5-7}\cmidrule(lr){8-10}\cmidrule(lr){11-13}
Task & \shortstack{Dense\\(Triton)} & FPv2 & BFLA & \shortstack{Dense\\(FA3)} & FPv2 & BFLA & \shortstack{Dense\\(FA3)} & FPv2 & BFLA & \shortstack{Dense\\(FA3)} & QSA & BFLA \\
\midrule
Retri.PassKey & 100.00 & 100.00 & 100.00 & 100.00 & 100.00 & 100.00 & 100.00 & 100.00 & 100.00 & 100.00 & 100.00 & 100.00 \\
Retri.Number & 100.00 & 100.00 & 100.00 & 100.00 & 100.00 & 100.00 & 100.00 & 100.00 & 100.00 & 100.00 & 100.00 & 100.00 \\
Retri.KV & 62.00 & 64.20 & 69.60 & 77.40 & 77.40 & 77.40 & 77.40 & 77.00 & 77.40 & 100.00 & 100.00 & 100.00 \\
Code.Run & 2.25 & 0.75 & 2.00 & 10.50 & 5.00 & 7.75 & 4.00 & 1.00 & 3.50 & 10.25 & 11.75 & 7.00 \\
Code.Debug & 40.10 & 38.58 & 41.12 & 77.41 & 74.87 & 77.66 & 78.43 & 73.60 & 76.90 & 80.96 & 85.79 & 80.71 \\
En.MC & 62.88 & 62.01 & 62.45 & 76.42 & 76.42 & 76.86 & 84.28 & 82.53 & 84.28 & 89.08 & 91.27 & 90.83 \\
En.QA & 26.55 & 27.09 & 26.51 & 49.15 & 45.63 & 48.79 & 47.96 & 42.92 & 47.23 & 49.52 & 53.66 & 50.83 \\
Zh.QA & 11.37 & 11.61 & 11.50 & 27.23 & 28.28 & 27.45 & 17.15 & 18.11 & 18.16 & 20.89 & 23.53 & 20.71 \\
En.Dialogue & 27.00 & 23.50 & 31.50 & 90.00 & 89.00 & 90.50 & 86.00 & 58.50 & 83.00 & 87.00 & 92.50 & 86.50 \\
Math.Find & 32.00 & 34.00 & 31.71 & 49.14 & 53.43 & 51.43 & 77.71 & 71.71 & 76.57 & 63.43 & 62.57 & 67.71 \\
En.Summary & 30.16 & 30.14 & 30.55 & 33.01 & 33.49 & 32.84 & 35.71 & 35.04 & 35.66 & 32.50 & 33.56 & 32.37 \\
Math.Calc & 0.0004 & 0.0005 & 0.0004 & 0.0003 & 0.0003 & 0.0003 & 0.0003 & 0.0003 & 0.0003 & 0.0003 & 0.0003 & 0.0003 \\
\midrule
Macro avg. & 41.19 & 40.99 & 42.25 & 57.52 & 56.96 & 57.56 & 59.05 & 55.03 & 58.56 & 61.14 & 62.89 & 61.39 \\
\bottomrule
\end{tabular}
\end{table*}

\subsection{Terminal Bench 2.1}
\label{sec:bfla_agentic_evaluation}

\paragraph{Deployment protocol.}
On Qwen3.8-Flash-Next, we compare dense FA3, QSA, and BFLA for prefill using the same FP8 weights, two GPUs with tensor parallelism, and native QSA decoding. All deployments enable thinking with \texttt{xhigh} reasoning effort and retain the default sampling settings: temperature~1.0, Top-$p=0.95$, and vocabulary-sampling Top-$k=20$. The context limit is 262,144 tokens (256K), shared by input and generated tokens. Evaluation concurrency is two, with a shared 16,384-token chunked-prefill budget per scheduler step. We follow the Terminal-Bench 2.1 methodology\footnote{\url{https://artificialanalysis.ai/methodology/intelligence-benchmarking\#terminal-bench-v2-1}} to run each task. We report pass@1 averaged over three repetitions of all 89 tasks.

\begin{table*}[htbp]
\centering
\small
\setlength{\tabcolsep}{4pt}
\caption{InfiniteBench: Mean request latencies (seconds) and overall speedups over dense baselines.}
\label{tab:bfla_infinitebench_runtime}
\begin{tabular}{llrrrr}
\toprule
 & & \multicolumn{2}{c}{Mean latency (s)} & \multicolumn{2}{c}{Overall speedup} \\
\cmidrule(lr){3-4}\cmidrule(lr){5-6}
Model & Method & TTFT & E2E & TTFT & E2E \\
\midrule
Gemma4-E4B-IT & Dense (Triton) & 12.983 & 13.764 & $1.000\times$ & $1.000\times$ \\
 & FlashPrefill V2 & 3.465 & 5.023 & $3.747\times$ & $2.740\times$ \\
 & BFLA (ours) & 3.324 & 4.272 & $3.906\times$ & $3.222\times$ \\
\midrule
Qwen3.6-27B & Dense (FA3) & 13.885 & 20.630 & $1.000\times$ & $1.000\times$ \\
 & FlashPrefill V2 & 10.151 & 17.031 & $1.368\times$ & $1.211\times$ \\
 & BFLA (ours) & 11.298 & 18.211 & $1.229\times$ & $1.133\times$ \\
\midrule
Qwen3.8-27B & Dense (FA3) & 13.884 & 20.390 & $1.000\times$ & $1.000\times$ \\
 & FlashPrefill V2 & 10.206 & 16.770 & $1.360\times$ & $1.216\times$ \\
 & BFLA (ours) & 11.166 & 17.945 & $1.243\times$ & $1.136\times$ \\
\midrule
Qwen3.8-Flash-Next & Dense (FA3) & 6.686 & 8.434 & $1.000\times$ & $1.000\times$ \\
 & QSA & 4.807 & 6.072 & $1.391\times$ & $1.389\times$ \\
 & BFLA (ours) & 4.983 & 6.459 & $1.342\times$ & $1.306\times$ \\
\bottomrule
\end{tabular}
\end{table*}

\paragraph{Task success and reasoning length.}
As shown in Table~\ref{tab:bfla_terminalbench_results}, BFLA achieves 86.89\% mean pass@1, compared with 85.77\% for QSA and 86.52\% for dense FA3, improvements of 1.12 and 0.37 percentage points, respectively. Even compared to the official score of 86.10\% reported by Artificial Analysis\footnote{https://artificialanalysis.ai/evaluations/terminalbench-2-1}, BFLA still matches performance. These results support preserving agentic task performance with training-free sparse prefill in this setting. Separately, we examine the retained trajectories by retokenizing logged main-agent and subagent thinking text with the same model tokenizer. BFLA's mean per-task thinking length is 1.7\% lower than QSA's and 2.7\% higher than dense FA3's.

\begin{table}[htbp]
\centering
\small
\caption{Terminal Bench 2.1 on Qwen3.8-Flash-Next. Pass@1 is averaged over three repetitions of 89 tasks.}
\label{tab:bfla_terminalbench_results}
\begin{tabular}{lrr}
\toprule
Prefill method & Mean pass@1 (\%) & Mean thinking tokens \\
\midrule
Dense (FA3) & 86.52 & 34875 \\
QSA         & 85.77 & 36423 \\
BFLA        & 86.89 & 35810 \\
\bottomrule
\end{tabular}
\end{table}

\subsection{Operator-Level Prefill Efficiency}
\label{sec:bfla_operator_efficiency}
\begin{figure*}[htbp]
    \centering
    \begin{minipage}[t]{\textwidth}
        \centering
        \includegraphics[width=0.8\linewidth]
        {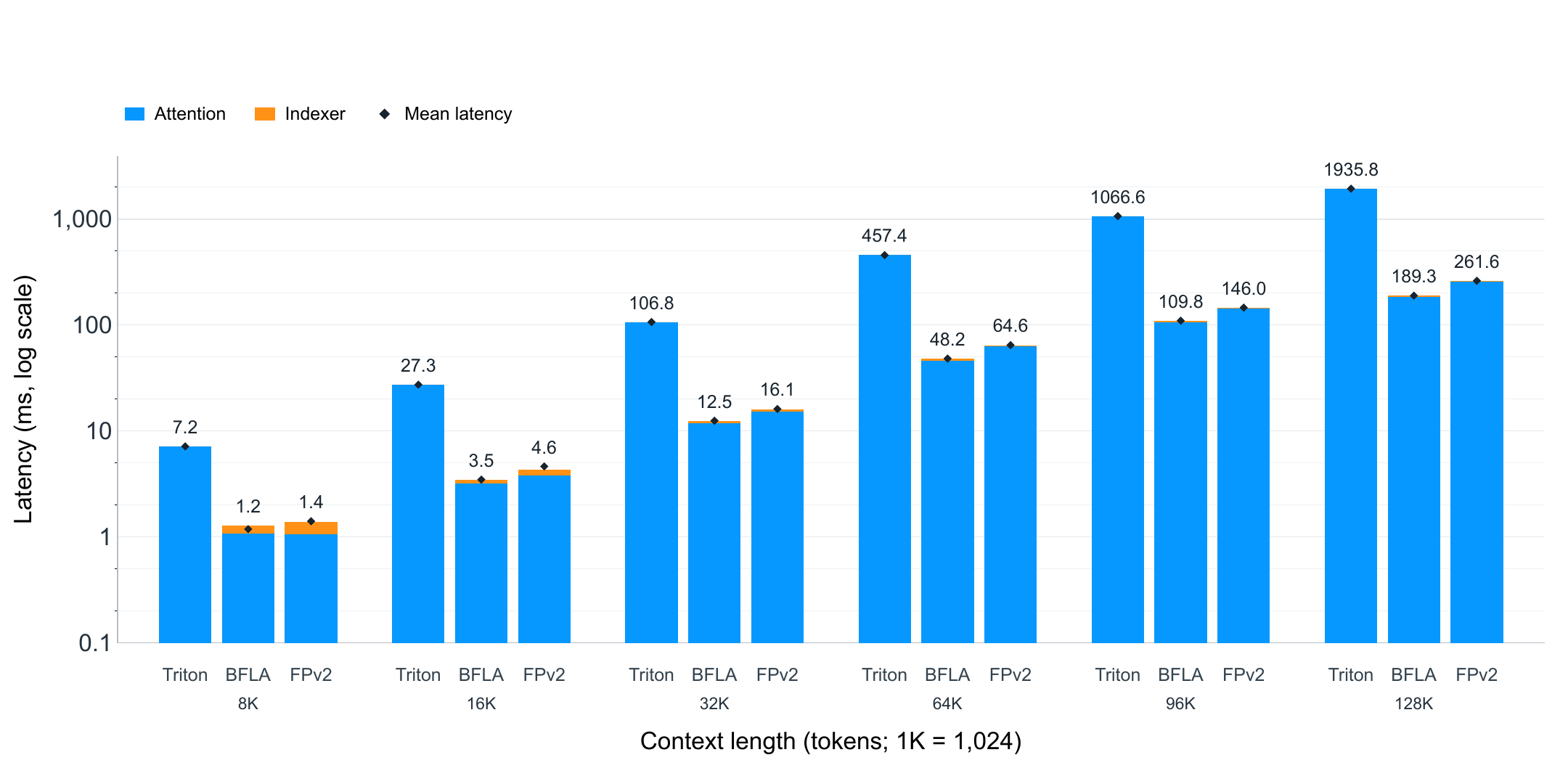}
        \par(a) Gemma4-E4B-IT
    \end{minipage}
    \begin{minipage}[t]{\textwidth}
        \centering
        \includegraphics[width=0.8\linewidth]
        {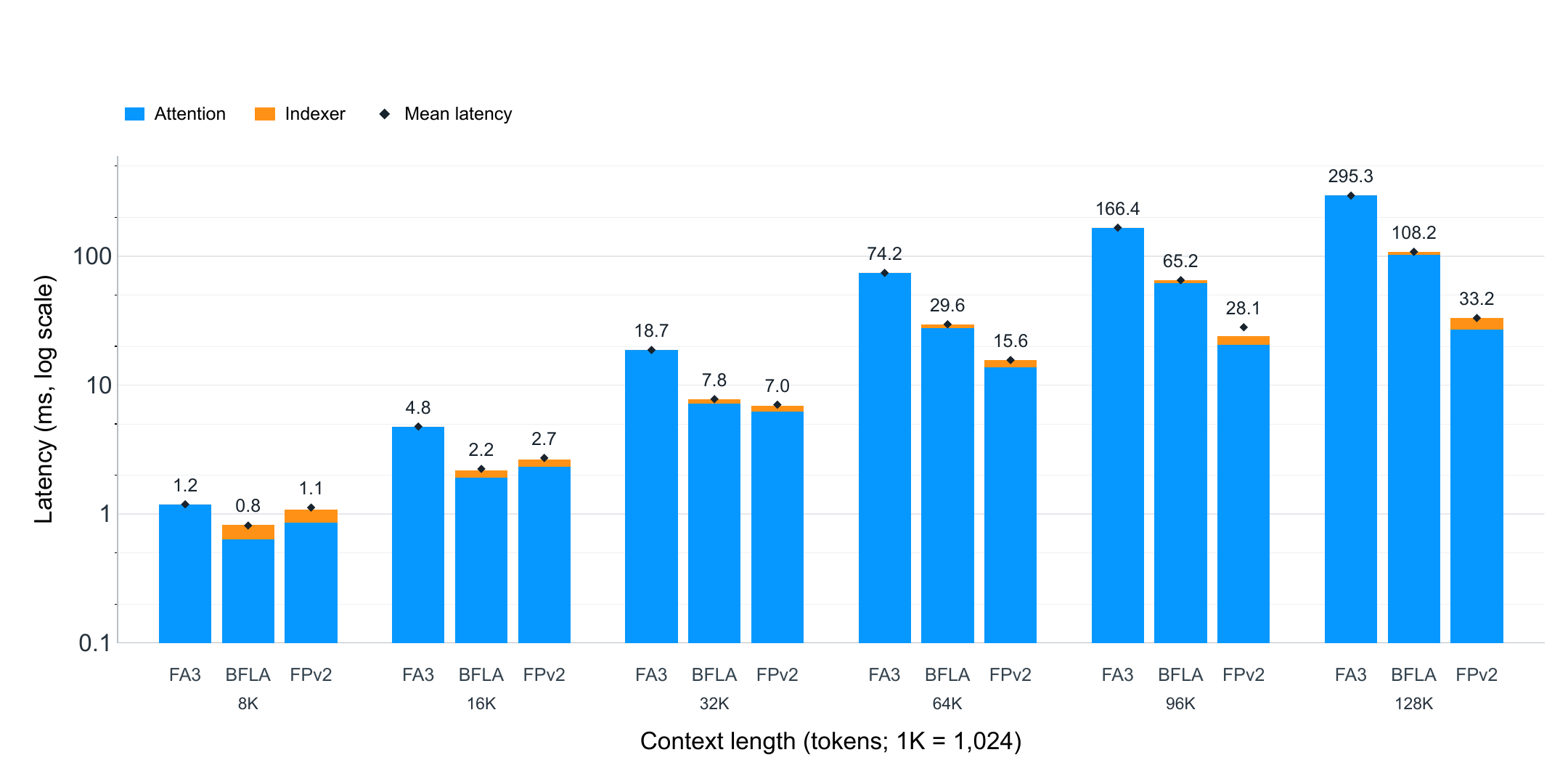}
        \par(b) Qwen3.8-27B
    \end{minipage}
    \begin{minipage}[t]{\textwidth}
        \centering
        \includegraphics[width=0.8\linewidth]
        {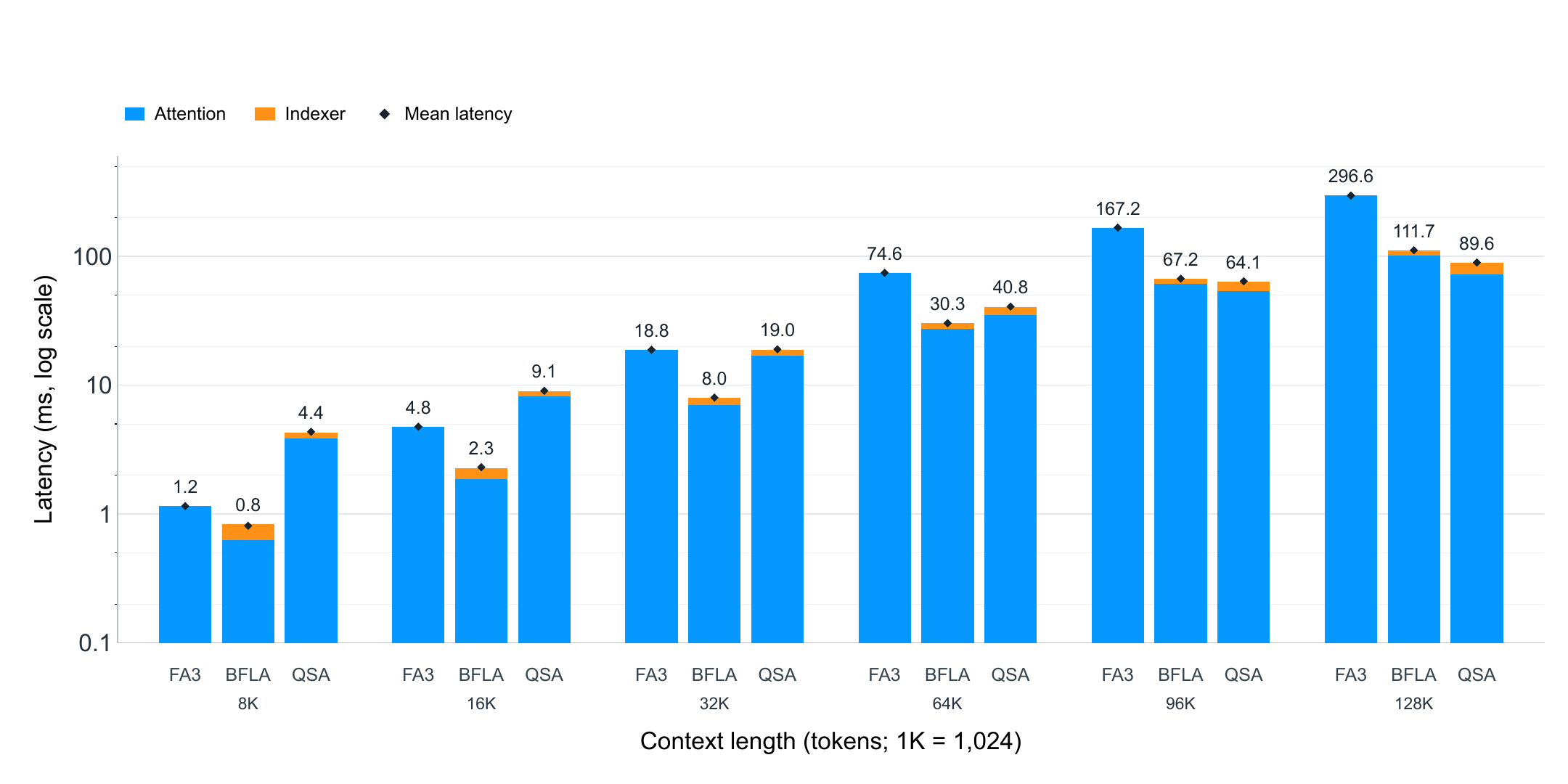}
        \par(c) Qwen3.8-Flash-Next
    \end{minipage}
    \caption{Prefill-pipeline latency on an H200 GPU at batch size one. Results average 300 measurements after 30 warm-up runs and include indexing, preprocessing, selection, and attention. FPv2 denotes FlashPrefill V2.}
    \label{fig:bfla_operator_latency}
\end{figure*}

Figure~\ref{fig:bfla_operator_latency} compares the prefill pipeline using matched model Q/K/V on an H200 GPU. BFLA achieves speedups of $6.00$--$10.23\times$ on Gemma, $1.50$--$2.73\times$ on Qwen3.8-27B, and $1.50$--$2.66\times$ on Qwen3.8-Flash-Next relative to the corresponding dense implementation. It is faster than FlashPrefill V2 throughout Gemma's tested range and at 8K--16K on Qwen3.8-27B. On Qwen3.8-Flash-Next, BFLA is faster than QSA at 8K--64K, with the ordering reversed at 96K--128K. The indexer of BFLA is faster than FlashPrefill V2 and QSA.

\section{Conclusions}
We present BFLA, a training-free sparse prefill method that jointly designs compact block indexing, query-tile sharing, and Ratio-$K$ selection with efficient sparse execution. BFLA integrates with vLLM while preserving native decoding, without retraining or per-model retuning. Across four models and three long-context benchmarks, it achieves up to approximately $4\times$ TTFT and $3.72\times$ end-to-end speedups over dense-prefill baselines, while matching or even outperforming state-of-the-art dense/sparse baselines. Comparisons with FlashPrefill V2 and QSA demonstrate competitive quality--latency trade-offs, with benefits varying across models and tasks. In agentic tasks, BFLA still achieves state-of-the-art performance. These findings highlight the value of BFLA for practical long-context acceleration.

\bibliographystyle{IEEEtran}
\bibliography{references}

\clearpage
\appendix

\section{Selection-Policy Details}
\label{app:bfla_policy_details}

This section specifies the boundary conventions used by the selection policy.

\paragraph{Query intervals and tile membership.}
Consider $n_{\mathrm{seg}}$ consecutive query tokens starting at absolute position $i_0$. Within each KV-head group, order query vectors first by token and then by query head. For query position $i$ and head $h$, the row number is $(i-i_0)G+(h\bmod G)$, so its tile index is
\[
 m(i,h)=\left\lfloor\frac{(i-i_0)G+(h\bmod G)}{M}\right\rfloor.
\]
For a nonempty tile $m$, the first and last token positions represented by its row set $\mathcal R_m$ are
\[
 f_m=i_0+\left\lfloor\frac{\min\mathcal R_m}{G}\right\rfloor,
 \qquad
 e_m=i_0+\left\lfloor\frac{\max\mathcal R_m}{G}\right\rfloor.
\]
The tile covers $e_m-f_m+1$ distinct token positions, which equals $M/G$ only for a full tile when $G$ divides $M$. A boundary may split one token's query heads across adjacent tiles. Tile indices are local to the query interval; the request and interval are implicit parts of the support identifier.

\paragraph{Mandatory blocks.}
Let $n_{\mathrm{sink}}\ge0$ be the number of protected initial tokens, and let $b_{\mathrm{loc}}\ge1$ and $n_{\mathrm{loc}}\ge1$ be integer local-window alignment and count parameters. With candidate set $\mathcal C_m$,
\begin{equation}
 \begin{aligned}
 \ell_m&=b_{\mathrm{loc}}\max\{0,
          \lfloor f_m/b_{\mathrm{loc}}\rfloor-n_{\mathrm{loc}}+1\},\\
 \mathcal M_m&=\mathcal C_m\cap\bigl(
     \{j:j<\lceil n_{\mathrm{sink}}/b\rceil\}
     \cup\{j:j\ge\lfloor f_m/b\rfloor\}
     \cup\{j:j\ge\lfloor\ell_m/b\rfloor\}\bigr).
 \end{aligned}
 \label{eq:bfla_mandatory}
\end{equation}
The three terms retain sink blocks, blocks overlapping the query range, and the aligned local band. Set union counts overlaps once.

\paragraph{Protected full-prompt suffix.}
Tail protection uses a positive integer reference group size $M_{\mathrm{tail}}$ and a nonnegative integer group count $r_{\mathrm{tail}}$. Within each KV-head group, the full prompt contains $PG$ query vectors ordered by token position and then by head. Partition these vectors into consecutive reference groups of at most $M_{\mathrm{tail}}$ rows and protect the final $r_{\mathrm{tail}}$ groups, or all groups if fewer are available. The first protected token position is
\begin{equation}
 N_s(P)=
 \begin{cases}
  P, &r_{\mathrm{tail}}=0,\\
  0,&P\le K_{\min}\ \text{and}\ r_{\mathrm{tail}}>0,\\
  \left\lfloor\dfrac{M_{\mathrm{tail}}}{G}
    \max\left\{0,
      \left\lceil\dfrac{PG}{M_{\mathrm{tail}}}\right\rceil
      -r_{\mathrm{tail}}\right\}\right\rfloor,
      &P>K_{\min}\ \text{and}\ r_{\mathrm{tail}}>0.
 \end{cases}
 \label{eq:bfla_tail}
\end{equation}
Rounding down protects every head of the boundary token. The suffix contains $P-N_s(P)$ query positions, not additional KV blocks. The reference size $M_{\mathrm{tail}}$ is distinct from the selection-sharing size $M$, and the boundary is based on the complete prompt. When $r_{\mathrm{tail}}=0$, $N_s(P)=P$ and the suffix is empty; the $N\le K_{\min}$ full-attention rule still applies.

\paragraph{Block ordering.}
Selection uses a different block quota for each query tile. Equal indexer scores are resolved in favor of smaller logical block IDs; after selection, IDs are sorted in increasing order for traversal. This changes traversal order, not the selected set.

\section{Proofs and Supporting Analysis}
\label{app:bfla_theory}
\label{UpperANA}
This appendix proves the results in Section~\ref{sec:theory} and develops their implications for shared indexing and retention budgets. We analyze one prefill call for a fixed request and layer, using the same fixed, finite original Q/K/V for the dense and sparse computations in real arithmetic.

\paragraph{Notation.}
We inherit the notation of Section~\ref{sec:method}. Query positions satisfy $i\in\mathcal I$, while causal key positions range over $0\le t\le i$. Tile identifiers retain their request and query-interval identity. The current query positions governed by score-based selection are
\[
 \mathcal I_{\mathrm{sel}}
 =\{i\in\mathcal I:N>K_{\min},\ i<N_s(P)\}.
\]
All sparse outputs in this appendix refer to the exact real-arithmetic expression on the selected support; floating-point evaluation adds a separate numerical error. The vector $\mathbf o_{i,h}$ is the attention output before model-specific output gating and projection. We distinguish the analytical score distortion $\varepsilon^{\mathrm{rank}}_{i,h}$ from the RMS stabilizer $\epsilon$.

\subsection{Preserving Attention Mass Controls Local Error}
\label{app:bfla_truncation}

For $i\in\mathcal I$ and query head $h$, write the original logits and normalized dense attention weights as
\begin{equation}
 z_{i,h,t}=\sigma\mathbf q_{i,h}\mathbf k_{t,\kappa(h)}^\top,
 \qquad
 p_{i,h,t}
 =\frac{e^{z_{i,h,t}}}{\sum_{v=0}^{i}e^{z_{i,h,v}}},
 \quad 0\le t\le i.
 \label{eq:bfla_bound_dense_weights}
\end{equation}
For the nonempty retained support $\mathcal A_{i,h}$ in Eq.~\eqref{eq:bfla_support}, define
\begin{equation}
 \delta_{i,h}
 =\sum_{\substack{0\le t\le i\\t\notin\mathcal A_{i,h}}}p_{i,h,t},
 \qquad
 \widehat p_{i,h,t}
 =\frac{p_{i,h,t}\mathbf 1\{t\in\mathcal A_{i,h}\}}
        {1-\delta_{i,h}}.
 \label{eq:bfla_bound_omitted_mass}
\end{equation}
Write $\mathbf p_{i,h}=(p_{i,h,t})_{t=0}^{i}$ and $\widehat{\mathbf p}_{i,h}=(\widehat p_{i,h,t})_{t=0}^{i}$. Thus $\delta_{i,h}$ is the fraction of original dense attention mass omitted by the retained support. Finite logits and nonempty retained support imply $0\le\delta_{i,h}<1$.

\begin{bflaboundlemma}[Normalized support restriction]
\label{lem:bfla_mass_identity}
Let $\mathbf o_{i,h}$ be the BFLA output and, when $\delta_{i,h}>0$, let $\mathbf o^{\mathrm{drop}}_{i,h}$ be the conditional mean of the values on the omitted support. When $\delta_{i,h}=0$, set $\mathbf o^{\mathrm{drop}}_{i,h}=\mathbf o_{i,h}$. Then
\begin{equation}
 \|\widehat{\mathbf p}_{i,h}-\mathbf p_{i,h}\|_1
 =2\delta_{i,h},
 \qquad
 \mathbf o^{\mathrm{dense}}_{i,h}-\mathbf o_{i,h}
 =\delta_{i,h}
  (\mathbf o^{\mathrm{drop}}_{i,h}-\mathbf o_{i,h}).
 \label{eq:bfla_bound_identity}
\end{equation}
Define the value diameter
\[
 D^V_{i,h}
 =\max_{0\le t,v\le i}
   \|\mathbf v_{t,\kappa(h)}-\mathbf v_{v,\kappa(h)}\|_2.
\]
Let $\mathbf O^{\mathrm{dense}},\mathbf O\in\mathbb R^{N_q\times H_q\times d}$ collect the outputs of the current call. Then
\begin{equation}
 \begin{aligned}
 \|\mathbf o^{\mathrm{dense}}_{i,h}-\mathbf o_{i,h}\|_2
 &\le D^V_{i,h}\delta_{i,h},\\
 \|\mathbf O^{\mathrm{dense}}-\mathbf O\|_F^2
 &\le
 \sum_{i\in\mathcal I}\sum_{h=0}^{H_q-1}
 (D^V_{i,h})^2\delta_{i,h}^2.
 \end{aligned}
 \label{eq:bfla_bound_output}
\end{equation}
\end{bflaboundlemma}

\begin{proof}
On the retained support,
\[
 \sum_{t\in\mathcal A_{i,h}}(\widehat p_{i,h,t}-p_{i,h,t})
 =\frac{\delta_{i,h}}{1-\delta_{i,h}}
   \sum_{t\in\mathcal A_{i,h}}p_{i,h,t}
 =\delta_{i,h}.
\]
On the omitted support, the absolute decrements also sum to $\delta_{i,h}$, proving the $L_1$ identity. Splitting the dense weighted sum gives
\[
 \mathbf o^{\mathrm{dense}}_{i,h}
 =(1-\delta_{i,h})\mathbf o_{i,h}
  +\delta_{i,h}\mathbf o^{\mathrm{drop}}_{i,h}.
\]
For $\delta_{i,h}>0$, the difference of conditional means can be written as
\[
 \mathbf o^{\mathrm{drop}}_{i,h}-\mathbf o_{i,h}
 =\sum_{\substack{0\le t\le i\\t\notin\mathcal A_{i,h}}}
  \sum_{v\in\mathcal A_{i,h}}
  \frac{p_{i,h,t}}{\delta_{i,h}}
  \frac{p_{i,h,v}}{1-\delta_{i,h}}
  (\mathbf v_{t,\kappa(h)}-\mathbf v_{v,\kappa(h)}).
\]
The coefficients are nonnegative and sum to one, so the triangle inequality gives the diameter bound. When $\delta_{i,h}=0$, the outputs coincide directly. The squared Frobenius norm is the sum of squared Euclidean row errors, yielding the final inequality.
\end{proof}

These normalized-truncation arguments follow established attention tail-mass analyses \citep{TopKTV,Twilight}. For a fixed row, fixed logits, and fixed support, suppress the row and head indices. The worst-case output error over values satisfying $\|\mathbf v_t\|_2\le R$, for $R\ge0$ and $d\ge1$, equals $2R\delta$. The upper bound follows from the lemma, and equality is attained by assigning opposite vectors of norm $R$ to retained and omitted positions. Minimizing omitted mass therefore minimizes this worst-case bound, but need not minimize the realized error for fixed model values.

\subsection{Retention Budget and Oracle Coverage}
\label{app:bfla_budget_oracle}

Consider a score-selected tile $m$ with the quota $k_m$ from Eq.~\eqref{eq:bfla_ratio_budget}. Define
\begin{equation}
 \begin{aligned}
 \mathcal F_m&=\mathcal C_m\setminus\mathcal M_m,
 &
 J_m^{\mathrm{mand}}&=|\mathcal M_m|,\\
 J_m^{\mathrm{free}}&=|\mathcal F_m|,
 &
 k_m^{\mathrm{free}}&=k_m-J_m^{\mathrm{mand}}.
 \end{aligned}
 \label{eq:bfla_bound_free_budget}
\end{equation}
Let $\mathcal T_m\subseteq\mathcal F_m$ contain the $k_m^{\mathrm{free}}$ blocks selected by their indexer scores $S_{m,j}$, so that $\mathcal S_m=\mathcal M_m\cup\mathcal T_m$. We assume the feasibility condition $0\le k_m^{\mathrm{free}}\le J_m^{\mathrm{free}}$.

If $j\in\mathcal F_m$, query-range protection implies $j<\lfloor f_m/b\rfloor$. Hence every $t\in\mathcal B_j$ satisfies
\[
 t\le(j+1)b-1<b\lfloor f_m/b\rfloor\le f_m\le i.
\]
Thus each free block is complete and causally visible to every row in the tile. Absolute positions make this valid for chunked prefill as well. Finite logits then give strictly positive mass on every free block.

For a row $(i,h)$ assigned to tile $m=m(i,h)$, define
\begin{equation}
 \begin{aligned}
 W_{i,h,j}
 &=\sum_{\substack{t\in\mathcal B_j\\t\le i}}e^{z_{i,h,t}},\\
 \beta_{i,h}
 &=\frac{\sum_{j\in\mathcal M_m}W_{i,h,j}}
         {\sum_{t=0}^{i}e^{z_{i,h,t}}},\\
 \zeta_{i,h,j}
 &=\log W_{i,h,j},
 \qquad
 p^{\mathrm{free}}_{i,h,j}
 =\frac{W_{i,h,j}}{\sum_{v\in\mathcal F_m}W_{i,h,v}},
 \quad j\in\mathcal F_m.
 \end{aligned}
 \label{eq:bfla_bound_block_mass}
\end{equation}
Here $W_{i,h,j}$ is defined for every $j\in\mathcal C_m$; $\zeta_{i,h,j}$ and $p^{\mathrm{free}}_{i,h,j}$ are used only for $j\in\mathcal F_m$ when that set is nonempty. In that case its total mass is positive. The quantity $\beta_{i,h}$ is the original attention mass protected by mandatory blocks, and $p^{\mathrm{free}}_{i,h,j}$ is the conditional mass distribution over the remaining blocks.

For nonempty $\mathcal F_m$, let $p^{\mathrm{free}}_{i,h,(s)}$ denote these conditional masses in decreasing order. For $0\le k\le J_m^{\mathrm{free}}$, define the oracle tail
\begin{equation}
 \tau_{i,h}(k)
 =\sum_{s=k+1}^{J_m^{\mathrm{free}}}
   p^{\mathrm{free}}_{i,h,(s)}.
 \label{eq:bfla_bound_oracle_tail}
\end{equation}
Choosing the $k$ largest free-block masses minimizes omitted mass: exchanging a selected lower-mass block for an omitted higher-mass block cannot decrease retained mass. Since the nonempty causal intersections of candidate blocks partition $\{0,\ldots,i\}$ and every mandatory block is retained, any omitted token belongs to a block in $\mathcal F_m\setminus\mathcal T_m$. For nonempty $\mathcal F_m$, this gives the exact identity
\begin{equation}
 \delta_{i,h}
 =(1-\beta_{i,h})
   \sum_{j\in\mathcal F_m\setminus\mathcal T_m}
   p^{\mathrm{free}}_{i,h,j}.
 \label{eq:bfla_bound_residual_identity}
\end{equation}
If $J_m^{\mathrm{free}}=0$ or $k_m^{\mathrm{free}}=J_m^{\mathrm{free}}$, the omitted mass is zero. If $k_m^{\mathrm{free}}=0<J_m^{\mathrm{free}}$, it equals $1-\beta_{i,h}$. Current queries receiving full causal support also have zero local truncation error on fixed Q/K/V.

\subsection{Additional Retention Tolerates Imperfect Indexing}
\label{app:bfla_surrogate_bound}

For nonempty $\mathcal F_m$ and finite indexer scores, define the discrepancy between original block scores and shared indexer scores up to a common additive shift by
\begin{equation}
 \begin{aligned}
 \varepsilon^{\mathrm{rank}}_{i,h}
 &=\inf_{a_{\mathrm{shift}}\in\mathbb R}
   \max_{j\in\mathcal F_m}
   |\zeta_{i,h,j}-S_{m,j}-a_{\mathrm{shift}}|\\
 &=\frac12\left[
   \max_{j\in\mathcal F_m}(\zeta_{i,h,j}-S_{m,j})
   -\min_{j\in\mathcal F_m}(\zeta_{i,h,j}-S_{m,j})
   \right].
 \end{aligned}
 \label{eq:bfla_bound_score_distortion}
\end{equation}
To see the second equality, let $d_j=\zeta_{i,h,j}-S_{m,j}$ on $\mathcal F_m$. Any center has maximum distance at least half the range of the $d_j$, while their extreme-value midpoint attains that distance. Thus the infimum is attained. This measures differences in score values and can be positive even when the block ordering is correct. A positive rescaling $S_{m,j}\mapsto\lambda S_{m,j}$, with $\lambda>0$, preserves Top-$k$ selection. The following bounds remain valid after recomputing the discrepancy at that scale and replacing the selection gap $\Delta_m$ by $\lambda\Delta_m$.

\begin{bflaboundtheorem}[Mandatory-aware indexer--budget error bound]
\label{thm:bfla_ratio_error}
Suppose $1\le k_m^{\mathrm{free}}<J_m^{\mathrm{free}}$ and all indexer scores on $\mathcal F_m$ are finite. Define
\[
 \Delta_m
 =\min_{j\in\mathcal T_m}S_{m,j}
  -\max_{v\in\mathcal F_m\setminus\mathcal T_m}S_{m,v}
 \ge0.
\]
Then
\begin{equation}
 \delta_{i,h}
 \le(1-\beta_{i,h})
 \min\left\{
 e^{2\varepsilon^{\mathrm{rank}}_{i,h}}
 \tau_{i,h}(k_m^{\mathrm{free}}),
 \;
 \frac{
 (J_m^{\mathrm{free}}-k_m^{\mathrm{free}})
 e^{2\varepsilon^{\mathrm{rank}}_{i,h}-\Delta_m}
 }{
 k_m^{\mathrm{free}}
 +(J_m^{\mathrm{free}}-k_m^{\mathrm{free}})
 e^{2\varepsilon^{\mathrm{rank}}_{i,h}-\Delta_m}
 }
 \right\}.
 \label{eq:bfla_bound_main}
\end{equation}
Multiplying the right-hand side by $D^V_{i,h}$ bounds the row-output error.
\end{bflaboundtheorem}

\begin{proof}
Fix the row and tile. Within this proof, write $k_f=k_m^{\mathrm{free}}$, $J_f=J_m^{\mathrm{free}}$, $\varepsilon=\varepsilon^{\mathrm{rank}}_{i,h}$, and $\Delta=\Delta_m$, and suppress row and tile subscripts on scores and sets. Let $\mathcal T^\star$ be a true top-$k_f$ set on $\mathcal F$. The sets $\mathcal T^\star\setminus\mathcal T$ and $\mathcal T\setminus\mathcal T^\star$ have equal size. Pair their elements as $(j_-,j_+)$, where $j_-$ was omitted by BFLA and $j_+$ was selected. At the minimizing shift, write $\zeta_j=S_j+a_{\mathrm{shift}}+e_j$ with $|e_j|\le\varepsilon$. Since $S_{j_+}\ge S_{j_-}$,
\[
 \zeta_{j_-}-\zeta_{j_+}\le2\varepsilon,
 \qquad
 W_{j_-}\le e^{2\varepsilon}W_{j_+}.
\]
The BFLA-omitted set consists of $\mathcal T^\star\setminus\mathcal T$ and the blocks omitted by both selectors; the oracle-omitted set similarly consists of $\mathcal T\setminus\mathcal T^\star$ and the same common remainder. Summing the paired inequalities and using $e^{2\varepsilon}\ge1$ on that remainder gives
\[
 \sum_{j\in\mathcal F\setminus\mathcal T}W_j
 \le e^{2\varepsilon}\sum_{j\in\mathcal F\setminus\mathcal T^\star}W_j.
\]
Division by the positive total free-block mass proves the oracle-tail estimate.

For the second bound, every selected free block $j_+$ and omitted free block $j_-$ satisfy $\zeta_{j_+}-\zeta_{j_-}\ge\Delta-2\varepsilon$. Let $A=\sum_{j\in\mathcal F\setminus\mathcal T}W_j$ and $B=\sum_{j\in\mathcal T}W_j>0$. Summing $W_{j_-}\le e^{2\varepsilon-\Delta}W_{j_+}$ over all $(J_f-k_f)k_f$ selected--omitted pairs gives $k_f A\le(J_f-k_f)e^{2\varepsilon-\Delta}B$, and therefore
\[
 \frac{\sum_{j\in\mathcal F\setminus\mathcal T}W_j}
      {\sum_{j\in\mathcal T}W_j}
 \le\frac{J_f-k_f}{k_f}e^{2\varepsilon-\Delta}.
\]
Since $A/(A+B)=(A/B)/(1+A/B)$ and $x/(1+x)$ is increasing for $x\ge0$, this proves the second estimate. Multiply both estimates by $1-\beta_{i,h}$ using Eq.~\eqref{eq:bfla_bound_residual_identity}. Since both hold, their minimum is valid.
\end{proof}

Theorem~\ref{thm:bfla_indexer_budget} follows from the first term of Eq.~\eqref{eq:bfla_bound_main}, the trivial residual-mass bound $\delta_{i,h}\le1-\beta_{i,h}$, and Lemma~\ref{lem:bfla_mass_identity}. The next result makes the coverage provided by additional retention explicit.

\begin{bflaboundcorollary}[Oversampling protects a critical oracle set]
\label{cor:bfla_oversampling}
Under the same finite-score assumptions, let $\zeta_{i,h,(v)}$ be the original log masses of the free blocks in decreasing order, with fixed tie-breaking. For an integer $s$ satisfying $1\le s\le k_m^{\mathrm{free}}<J_m^{\mathrm{free}}$, if
\begin{equation}
 \zeta_{i,h,(s)}
 -\zeta_{i,h,(k_m^{\mathrm{free}}+1)}
 >2\varepsilon^{\mathrm{rank}}_{i,h},
 \label{eq:bfla_bound_oversampling_gap}
\end{equation}
then BFLA's free selection contains every true top-$s$ free block, and
\[
 \delta_{i,h}
 \le(1-\beta_{i,h})\tau_{i,h}(s).
\]
\end{bflaboundcorollary}

\begin{proof}
If a true top-$s$ block $j_-$ were omitted, the $k_m^{\mathrm{free}}$ selected free blocks would include a block $j_+$ outside the true top-$k_m^{\mathrm{free}}$ set. The assumed gap gives $\zeta_{j_-}-\zeta_{j_+}>2\varepsilon^{\mathrm{rank}}_{i,h}$, whereas the shift-centered discrepancy implies
\[
 S_{m,j_-}-S_{m,j_+}
 \ge \zeta_{j_-}-\zeta_{j_+}-2\varepsilon^{\mathrm{rank}}_{i,h}>0,
\]
contradicting Top-$k$ selection. Hence all true top-$s$ free blocks are retained, and the remaining omitted conditional mass is at most $\tau_{i,h}(s)$.
\end{proof}

Hold the original Q/K/V, candidates, mandatory blocks, and indexer scores fixed. For a fixed $s$, increasing $k_m^{\mathrm{free}}$ moves the competing block further down the true ranking and weakens the sufficient condition in Eq.~\eqref{eq:bfla_bound_oversampling_gap}. The indexer only needs to place the important blocks inside the retained set. Their internal ordering does not affect this coverage guarantee.

\paragraph{Comparing indexer--budget configurations.}
For fixed original Q/K/V, candidates, and mandatory blocks, the factors $D^V_{i,h}$ and $1-\beta_{i,h}$ are common to both configurations. Comparing the tail-based bounds therefore gives the sufficient condition in Eq.~\eqref{eq:bfla_main_compensation}. This condition orders the bounds on local output error. Changing query tiling or block size can also change the candidate sets, mandatory mass, and oracle tail, so such configurations must be bounded separately rather than compared through Eq.~\eqref{eq:bfla_main_compensation} alone.

\subsection{How Shared Indexing Enters the Trade-off}
\label{app:bfla_distortion}

On the nonempty coordinate set $\mathcal F_m$, define the shift-invariant seminorm
\[
 \|\mathbf x\|_{\mathrm{osc}}
 =\frac12\left(
   \max_{j\in\mathcal F_m}x_j
   -\min_{j\in\mathcal F_m}x_j
 \right).
\]
Let $\boldsymbol\zeta_{i,h}=(\zeta_{i,h,j})_{j\in\mathcal F_m}$ and $\mathbf S_m=(S_{m,j})_{j\in\mathcal F_m}$. For a nonempty score-selected tile with $\mathcal F_m\ne\varnothing$, restrict to its query interval and define
\[
 \begin{aligned}
 \mathcal U_m
 &=\{(i,h):i\in\mathcal I_{\mathrm{sel}},\ i_0\le i<i_0+n_{\mathrm{seg}},\
             0\le h<H_q,\ m(i,h)=m\},\\
 \overline{\boldsymbol\zeta}_m
 &=\frac1{|\mathcal U_m|}
   \sum_{(i,h)\in\mathcal U_m}\boldsymbol\zeta_{i,h}.
 \end{aligned}
\]
The set $\mathcal U_m$ includes all original query/head rows using the shared support, across KV-head groups. Their mean log-mass vector serves as an analytical reference.

By Eq.~\eqref{eq:bfla_bound_score_distortion}, $\varepsilon^{\mathrm{rank}}_{i,h}=\|\boldsymbol\zeta_{i,h}-\mathbf S_m\|_{\mathrm{osc}}$. The inequality $\max_j(x_j+y_j)-\min_j(x_j+y_j)\le(\max_jx_j-\min_jx_j)+(\max_jy_j-\min_jy_j)$ gives the seminorm triangle inequality and hence
\begin{equation}
 \varepsilon^{\mathrm{rank}}_{i,h}
 \le
 \underbrace{
 \|\boldsymbol\zeta_{i,h}
   -\overline{\boldsymbol\zeta}_m\|_{\mathrm{osc}}
 }_{\varepsilon^{\mathrm{share}}_{i,h}}
 +
 \underbrace{
 \|\overline{\boldsymbol\zeta}_m
   -\mathbf S_m\|_{\mathrm{osc}}
 }_{\varepsilon^{\mathrm{index}}_m}.
 \label{eq:bfla_bound_sharing}
\end{equation}
The first term captures differences in relevance across rows sharing a selection. The second captures the discrepancy between their average log masses and the compact indexer scores. This includes head/channel reduction, normalization, pooling, ReLU, and numerical representation.

Substituting this upper bound on score distortion into Eq.~\eqref{eq:bfla_bound_main} connects both effects to the retention budget. If Corollary~\ref{cor:bfla_oversampling} holds for every row in $\mathcal U_m$, the common support contains each row's top-$s$ free blocks. The condition must hold separately for every row whose coverage is claimed; in particular, the union of those rows' top-$s$ free blocks must fit within $k_m^{\mathrm{free}}$. Average agreement alone does not give per-row coverage.

\subsection{Retention Fraction and Tail Decay}
\label{app:bfla_ratio_interpretation}

For $1\le k_m^{\mathrm{free}}<J_m^{\mathrm{free}}$, define the retained fraction of free blocks as $\theta_m=k_m^{\mathrm{free}}/J_m^{\mathrm{free}}$. The second term of Eq.~\eqref{eq:bfla_bound_main} becomes
\begin{equation}
 \delta_{i,h}
 \le(1-\beta_{i,h})
 \frac{1-\theta_m}
      {1-\theta_m+
       \theta_m e^{\Delta_m-2\varepsilon^{\mathrm{rank}}_{i,h}}}.
 \label{eq:bfla_bound_ratio_envelope}
\end{equation}
The relevant fraction is
\[
 \theta_m
 =\frac{k_m-J_m^{\mathrm{mand}}}
        {|\mathcal C_m|-J_m^{\mathrm{mand}}},
\]
which is not generally equal to the configured ratio $\eta$. Its value follows from the quota in Eq.~\eqref{eq:bfla_ratio_budget} after mandatory retention and block rounding.

A positive effective gap $\Delta_m-2\varepsilon^{\mathrm{rank}}_{i,h}$ tightens the bound. With exact scores up to an additive shift and zero selection gap, Eq.~\eqref{eq:bfla_bound_ratio_envelope} reduces to $(1-\beta_{i,h})(1-\theta_m)$. This is sharp for uniform residual block mass, showing that a fixed retained fraction alone does not guarantee small error.

A stronger decay statement requires an assumption on attention concentration. Suppose
\[
 p^{\mathrm{free}}_{i,h,(s)}\le Cs^{-\nu},
 \qquad
 1\le s\le J_m^{\mathrm{free}},
 \qquad C>0,\quad \nu>1.
\]
Then the oracle tail for $1\le k\le J_m^{\mathrm{free}}$ satisfies the first inequality below. If additionally $1\le k_m^{\mathrm{free}}\le J_m^{\mathrm{free}}$ and the indexer scores are finite, the actual row error satisfies the second:
\begin{equation}
 \begin{aligned}
 \tau_{i,h}(k)
 &\le\frac{C}{\nu-1}k^{1-\nu},\\
 \|\mathbf o^{\mathrm{dense}}_{i,h}-\mathbf o_{i,h}\|_2
 &\le D^V_{i,h}(1-\beta_{i,h})
       e^{2\varepsilon^{\mathrm{rank}}_{i,h}}
       \frac{C}{\nu-1}
       \bigl(k_m^{\mathrm{free}}\bigr)^{1-\nu}.
 \end{aligned}
 \label{eq:bfla_bound_power_tail}
\end{equation}
Indeed,
\[
 \tau_{i,h}(k)\le C\sum_{s=k+1}^{\infty}s^{-\nu}
 \le C\int_k^\infty x^{-\nu}\,dx
 =\frac{C}{\nu-1}k^{1-\nu}.
\]
Substitution into the tail-based bound proves the second inequality. Full free-block retention has zero error directly. The case $k_m^{\mathrm{free}}=0$ uses the boundary identity in Appendix~\ref{app:bfla_budget_oracle}, not a negative power of zero.

Along a family of inputs for which the free budget grows, a fixed $\nu>1$ and uniform bounds on $C$, $\varepsilon^{\mathrm{rank}}_{i,h}$, and $D^V_{i,h}$ yield a decaying upper envelope. Once the retention quota saturates, the policy alone provides no further improvement in this tail bound. For fixed scores, candidates, mandatory blocks, and tie-breaking, increasing the quota yields nested supports and nonincreasing omitted mass, but need not produce monotone realized output error or downstream task accuracy.

\subsection{Supporting Bounds and Scope}
\label{app:bfla_bound_scope}

The main bound applies to the scores used for selection. The following results describe how block pooling and FP8 reconstruction can change those scores.

\paragraph{Block pooling and within-block variation.}
For a complete block $\mathcal B_j$ of size $b$, let all sums and extrema below range over $t\in\mathcal B_j$. Define
\[
 x_t
 =\frac{\boldsymbol\phi_i^Q(\boldsymbol\phi_t^K)^\top}
        {\sqrt{d_I}},
 \qquad
 y_t=[x_t]_+,
 \qquad
 u^0_{i,j}
 =\left[\frac1b\sum_t x_t\right]_+.
\]
Here $u^0_{i,j}$ is the unquantized block score obtained by pooling key features before applying ReLU. Define the token-based reference and the within-block variation terms as
\[
 \widetilde u_{i,j}
 =\log\left(\frac1b\sum_t e^{y_t}\right),
 \qquad
 \xi_{i,j}
 =\frac1b\sum_t y_t-u^0_{i,j},
 \qquad
 \omega_{i,j}
 =\max_t y_t-\min_t y_t.
\]
Then
\begin{equation}
 0\le\widetilde u_{i,j}-u^0_{i,j}
 \le\xi_{i,j}+\frac{\omega_{i,j}^2}{8}.
 \label{eq:bfla_bound_pooling}
\end{equation}

Let $\bar y=b^{-1}\sum_t y_t$ and $F(a)=\log(b^{-1}\sum_t e^{a(y_t-\bar y)})$. Under weights proportional to $e^{a(y_t-\bar y)}$, $F''(a)$ is the variance of $y_t$. Any distribution on $[y_{\min},y_{\max}]$ has variance at most $(y_{\max}-y_{\min})^2/4$: bound the mean squared distance to the interval midpoint, which is an upper bound on the variance. Thus $0\le F''(a)\le\omega_{i,j}^2/4$. Since $F(0)=F'(0)=0$,
\[
 F(1)=\int_0^1(1-a)F''(a)\,da
 \in[0,\omega_{i,j}^2/8].
\]
Convexity of ReLU gives $\xi_{i,j}\ge0$, and $\widetilde u_{i,j}-u^0_{i,j}=F(1)+\xi_{i,j}$ proves the claim. This is an index-space pooling bound, not a comparison with original per-head logits.

\paragraph{FP8 key-feature reconstruction.}
Let $\mathbf e_j^{\mathrm{idx}}=a_j\widehat{\boldsymbol\mu}_j-\boldsymbol\mu_j$ be the block key-feature reconstruction error. For any nonempty block, define its unquantized score as $u^0_{i,j}=[\boldsymbol\phi_i^Q\boldsymbol\mu_j^\top]_+/\sqrt{d_I}$. Holding the other operands fixed, this score and Eq.~\eqref{eq:bfla_shared_score} satisfy
\begin{equation}
 |u_{i,j}-u^0_{i,j}|
 \le
 \frac{\|\boldsymbol\phi_i^Q\|_2}{\sqrt{d_I}}
 \|\mathbf e_j^{\mathrm{idx}}\|_2.
 \label{eq:bfla_bound_quantization}
\end{equation}
Because $a_j>0$, positive homogeneity of ReLU rewrites the quantized score as $[\boldsymbol\phi_i^Q(a_j\widehat{\boldsymbol\mu}_j)^\top]_+/\sqrt{d_I}$. ReLU is 1-Lipschitz, and Cauchy--Schwarz then yields the bound. If two vectors of row scores differ by at most $c_0\ge0$ entrywise, their exponential sums differ by factors between $e^{-c_0}$ and $e^{c_0}$; their log-sum-exp values therefore differ by at most $c_0$. The shared aggregation does not amplify the maximum perturbation. The bound isolates block key-feature reconstruction error. Query-feature rounding and other arithmetic errors contribute separately.

\paragraph{Limits of deterministic indexing.}
When $c=d/d_I\ge2$, the channel-reduction map in Eq.~\eqref{eq:bfla_channel_map} has a nontrivial nullspace. In particular, $(1,-1,0,\ldots,0)$ is mapped to zero. This applies to $d_I=128$ with $d=256$ or $512$.

Let $\mathbf u=(1,-1,0,\ldots,0)$. Head normalization rescales multiples of $\mathbf u$ but does not move them out of the reduction map's nullspace; subsequent normalization and the Hadamard transform still give zero. For a nonempty omitted free block, set all relevant queries along $\mathbf u$, set its keys to $A\mathbf u$ with $A>0$, and set other keys to zero. Index scores remain tied and their fixed tie-breaking unchanged, while that block's original logits diverge as $A\to\infty$. Its dense attention mass approaches one. Deterministic indexing, FP8 precision, and a fixed retention ratio therefore do not by themselves guarantee small score distortion.

\paragraph{A conditional diagnostic from indexer scores.}
For nonempty $\mathcal F_m$ and finite indexer scores, normalize the free-block scores and define their omitted mass by
\[
 p^{\mathrm{idx}}_{m,j}
 =\frac{e^{S_{m,j}}}
        {\sum_{v\in\mathcal F_m}e^{S_{m,v}}},
 \qquad
 \tau^{\mathrm{idx}}_m
 =\sum_{j\in\mathcal F_m\setminus\mathcal T_m}
   p^{\mathrm{idx}}_{m,j}.
\]
These are analytical quantities, distinct from the original token weights $p_{i,h,t}$ and retained-support weights $\widehat p_{i,h,t}$. The score-distortion comparison gives
\begin{equation}
 \delta_{i,h}
 \le(1-\beta_{i,h})
 \frac{
 e^{2\varepsilon^{\mathrm{rank}}_{i,h}}\tau^{\mathrm{idx}}_m
 }{
 1-\tau^{\mathrm{idx}}_m
 +e^{2\varepsilon^{\mathrm{rank}}_{i,h}}\tau^{\mathrm{idx}}_m
 }.
 \label{eq:bfla_bound_proxy_tail}
\end{equation}
At a minimizing shift, each original weight $W_{i,h,j}$ lies between $e^{a_{\mathrm{shift}}-\varepsilon^{\mathrm{rank}}_{i,h}}e^{S_{m,j}}$ and $e^{a_{\mathrm{shift}}+\varepsilon^{\mathrm{rank}}_{i,h}}e^{S_{m,j}}$. For partial retention, summing the upper estimate over omitted blocks and the lower estimate over retained blocks bounds the true omitted-to-retained ratio by
\[
 e^{2\varepsilon^{\mathrm{rank}}_{i,h}}
 \frac{\tau^{\mathrm{idx}}_m}{1-\tau^{\mathrm{idx}}_m}.
\]
Converting this ratio using $x\mapsto x/(1+x)$ and multiplying by $1-\beta_{i,h}$ proves the claim. Empty and full selections satisfy the displayed bound directly with $\tau^{\mathrm{idx}}_m=1$ and $0$, respectively; no division by zero is used.

Interpreting retained indexer-score mass as a guarantee on original attention coverage requires a justified bound on score distortion. A budget rule based on this diagnostic would therefore require additional information beyond the current Ratio-$K$ policy.

\paragraph{Scope and empirical interpretation.}
The bounds concern the scores actually used for selection and apply whether the indexer key state is rebuilt or reused. Their discrepancy from the current original block scores is measured by $\varepsilon^{\mathrm{rank}}_{i,h}$. The relationship between indexer feature dimension and the resulting distortion remains an empirical question.

On sampled fixed Q/K/V, the mandatory mass $\beta_{i,h}$, oracle tail $\tau_{i,h}(k_m^{\mathrm{free}})$, actual omitted mass, and score distortion could distinguish budget limitations from indexing or sharing errors. The experiments report the task quality and runtime achieved by the complete configuration.

Full-support queries have zero local truncation error on the same operands, but their inputs and cached history can already differ from those of a separate dense execution. The bounds describe the local indexer--budget trade-off, while the experiments assess the behavior of the complete inference pipeline.

\section{Ablation Studies}
\label{sec:bfla_ablation}

\paragraph{Experimental setup.}
We vary one parameter at a time on LongBench-v2 using Gemma4-E4B-IT and Qwen3.8-27B. The default configuration is $\eta=1/3$, $M=128$, $K_{\min}=2048$, $K_{\max}=32768$, $b=64$, $d_I=128$, and $r_{\mathrm{tail}}=8$, with FP8 indexer key-state storage. All other settings follow the default configuration in Section~\ref{sec:bfla_experimental_setup}. Each configuration is evaluated over three complete runs on one H200 GPU at batch size one.

\begin{table*}[tbp]
\centering
\small
\setlength{\tabcolsep}{5pt}
\caption{Single-parameter LongBench-v2 ablations with a 2,048-token maximum output length. Acc. denotes accuracy. All values are three-run means. Unspecified parameters retain their default values.}
\label{tab:bfla_ablation}
\begin{tabular}{lrrrrrr}
\toprule
& \multicolumn{3}{c}{Gemma4-E4B-IT}
& \multicolumn{3}{c}{Qwen3.8-27B} \\
\cmidrule(lr){2-4}\cmidrule(lr){5-7}
Configuration
& Acc. (\%) & TTFT (s) & E2E (s)
& Acc. (\%) & TTFT (s) & E2E (s) \\
\midrule
\textbf{Default}
& 37.11 & 2.312 & 8.824
& 51.69 & 8.206 & 20.767 \\
\midrule
$\eta=1/2$
& 36.85 & 2.512 & 9.156
& 51.95 & 8.443 & 21.145 \\
$\eta=1/4$
& 34.46 & 2.129 & 8.802
& 50.23 & 7.970 & 20.978 \\
$\eta=1/5$
& 36.45 & 2.026 & 8.765
& 49.50 & 7.834 & 21.372 \\
\midrule
$M=256$
& 36.91 & 2.303 & 8.875
& 50.30 & 8.207 & 21.291 \\
$M=512$
& 36.32 & 2.290 & 8.928
& 50.03 & 8.161 & 21.382 \\
\midrule
$K_{\min}=1024$
& 36.85 & 2.314 & 8.914
& 49.97 & 8.195 & 20.962 \\
$K_{\min}=4096$
& 36.85 & 2.314 & 8.806
& 50.70 & 8.199 & 21.082 \\
$K_{\min}=8192$
& 37.77 & 2.329 & 8.889
& 51.03 & 8.231 & 20.758 \\
\midrule
$K_{\max}=8192$
& 37.31 & 1.856 & 8.516
& 49.70 & 7.608 & 20.799 \\
No upper selection limit $K_{\max}$
& 37.24 & 2.339 & 8.923
& 50.83 & 8.245 & 20.887 \\
\midrule
$b=128$
& 36.05 & 2.295 & 8.838
& 51.36 & 8.187 & 21.145 \\
$b=256$
& 35.72 & 2.287 & 8.961
& 49.24 & 8.151 & 21.070 \\
\midrule
$d_I=64$
& 35.65 & 2.305 & 8.873
& 51.09 & 8.205 & 20.861 \\
$d_I=256$
& 37.04 & 2.327 & 8.878
& 51.76 & 8.214 & 20.887 \\
BF16 indexer key state
& 36.91 & 2.300 & 8.884
& 50.70 & 8.179 & 20.952 \\
\midrule
$r_{\mathrm{tail}}=0$
& 37.44 & 2.290 & 8.861
& 51.36 & 8.222 & 21.232 \\
$r_{\mathrm{tail}}=4$
& 37.18 & 2.311 & 8.774
& 51.49 & 8.221 & 21.413 \\
$r_{\mathrm{tail}}=16$
& 37.97 & 2.318 & 8.917
& 51.23 & 8.234 & 20.829 \\
\bottomrule
\end{tabular}
\end{table*}

\paragraph{Experimental results.}
The ablations (Table~\ref{tab:bfla_ablation}) support a compact indexer without excessive compression or sharing. Reducing the indexer feature dimension from 128 to 64 lowers accuracy on both models, whereas increasing it to 256 yields little change. Replacing FP8 key-state storage with BF16 provides no accuracy gain. Larger pooling blocks and query-sharing groups also reduce accuracy on both models, with only minor changes in TTFT. These results support $d_I=128$, FP8 storage, $b=64$, and $M=128$: further compression or coarser selection sacrifices quality without a substantial request-level speed benefit, while wider features and higher precision offer limited additional value. Within the tested range, the retention budget has a greater effect on TTFT than the indexer dimension, storage precision, or sharing granularity. Smaller budgets provide larger speed gains, but their accuracy impact differs across models. Varying $r_{\mathrm{tail}}$ produces only small changes in accuracy and request latency on both models. We therefore retain the common default as a cross-model quality--efficiency compromise, rather than optimizing each model separately.

\end{document}